\documentclass[letterpaper,12pt]{article}
\usepackage{paper,math}
\pdfoutput=1
\renewcommand{\itshape}{\fontfamily{LinuxLibertineT-LF}\fontshape{it}\selectfont}
\renewcommand{\scshape}{\fontfamily{LinuxLibertineT-LF}\fontshape{sc}\selectfont}
\def\bib{paper.bib}
\def\pdf{figures.pdf}
\available{https://www.pascalmichaillat.org/11.html}
\hypersetup{pdftitle={Resolving New Keynesian Anomalies with Wealth in the Utility Function}}
\hypersetup{pdfauthor={Pascal Michaillat and Emmanuel Saez}}

\bmdefine{\M}{M}
\bmdefine{\v}{v}

\begin{document}

\title{Resolving New Keynesian Anomalies\\with Wealth in the Utility Function}
\author{Pascal Michaillat, Emmanuel Saez
\thanks{Michaillat: Brown University. Saez: University of California--Berkeley. We thank Sushant Acharya, Adrien Auclert, Gadi Barlevi, Marco Bassetto, Jess Benhabib, Florin Bilbiie, Jeffrey Campbell, Edouard Challe, Varanya Chaubey, John Cochrane, Behzad Diba, Gauti Eggertsson, Erik Eyster, Francois Gourio, Pete Klenow, Olivier Loisel, Neil Mehrotra, Emi Nakamura, Sam Schulhofer-Wohl, David Sraer, Jon Steinsson, Harald Uhlig, and Ivan Werning for helpful discussions and comments. This work was supported by the Institute for Advanced Study and the Berkeley Center for Equitable Growth.}}
\date{December 2019}

\begin{titlepage}\maketitle\begin{abstract}

At the zero lower bound, the New Keynesian model predicts that output and inflation collapse to implausibly low levels, and that government spending and forward guidance have implausibly large effects. To resolve these anomalies, we introduce wealth into the utility function; the justification is that wealth is a marker of social status, and people value status. Since people partly save to accrue social status, the Euler equation is modified. As a result, when the marginal utility of wealth is sufficiently large, the dynamical system representing the zero-lower-bound equilibrium transforms from a saddle to a source---which resolves all the anomalies.

\end{abstract}\end{titlepage}\section{Introduction}

A current issue in monetary economics is that the New Keynesian model makes several anomalous predictions when the zero lower bound on nominal interest rates (ZLB) is binding: implausibly large collapse of output and inflation \cp{EW04,E11,W12}; implausibly large effect of forward guidance \cp{DGP12,CFP15,C17}; and implausibly large effect of government spending \cp{CER09,W11,C17}.

Several papers have developed variants of the New Keynesian model that behave well at the ZLB \cp{G16,DL17,C18,B18,AD18}; but these variants are more complex than the standard model. In some cases the derivations are complicated by bounded rationality or heterogeneity. In other cases the dynamical system representing the equilibrium---normally composed of an Euler equation and a Phillips curve---includes additional differential equations that describe bank-reserve dynamics, price-level dynamics, or the evolution of the wealth distribution. Moreover, a good chunk of the analysis is conducted by numerical simulations. Hence, it is sometimes difficult to grasp the nature of the anomalies and their resolutions.

It may therefore be valuable to strip the logic to the bone. We do so using a New Keynesian model in which relative wealth enters the utility function. The justification for the assumption is that relative wealth is a marker of social status, and people value high social status. We deviate from the standard model only minimally: the derivations are the same; the equilibrium is described by a dynamical system composed of an Euler equation and a Phillips curve; the only difference is an extra term in the Euler equation. We also veer away from numerical simulations and establish our results with phase diagrams describing the dynamics of output and inflation given by the Euler-Phillips system. The model's simplicity and the phase diagrams allow us to gain new insights into the anomalies and their resolutions.\footnote{Our approach relates to the work of \ct{MS14}, \ct{OY12}, and \ct{M15}. By assuming wealth in the utility function, they obtain non-New-Keynesian models that behave well at the ZLB. But their results are not portable to the New Keynesian framework because they require strong forms of wage or price rigidity (exogenous wages, fixed inflation, or downward nominal wage rigidity). Our approach also relates to the work of \ct{F15} and \ct{CFJ17}, who build New Keynesian models with government bonds in the utility function. The bonds-in-the-utility assumption captures special features of government bonds relative to other assets, such as safety and liquidity \eg{KVJ12}. While their assumption and ours are conceptually different, they affect equilibrium conditions in a similar way. These papers use their assumption to generate risk-premium shocks (\name{F15}) and to alleviate the forward-guidance puzzle (\name{CFJ17}).}

Using the phase diagrams, we begin by depicting the anomalies in the standard New Keynesian model. First, we find that output and inflation collapse to unboundedly low levels when the ZLB episode is arbitrarily long-lasting. Second, we find that there is a duration of forward guidance above which any ZLB episode, irrespective of its duration, is transformed into a boom. Such boom is unbounded when the ZLB episode is arbitrarily long-lasting. Third, we find that there is an amount of government spending at which the government-spending multiplier becomes infinite when the ZLB episode is arbitrarily long-lasting. Furthermore, when government spending exceeds this amount, an arbitrarily long ZLB episode prompts an unbounded boom.

The phase diagrams also pinpoint the origin of the anomalies: they arise because the Euler-Phillips system is a saddle at the ZLB. In normal times, by contrast, the Euler-Phillips system is source, so there are no anomalies. In economic terms, the anomalies arise because household consumption (given by the Euler equation) responds too strongly to the real interest rate. Indeed, since the only motive for saving is future consumption, households are very forward-looking, and their response to interest rates is strong.

Once wealth enters the utility function, however, the Euler equation is ``discounted''---in the sense of \ct{MNS17}---which alters the properties of the Euler-Phillips system. People now save partly because they enjoy holding wealth; this is a present consideration, which does not require them to look into the future. As people are less forward-looking, their consumption responds less to interest rates; this creates discounting. 

With enough marginal utility of wealth, the discounting is strong enough to transform the Euler-Phillips system from a saddle to a source at the ZLB and thus eliminate all the anomalies. First, output and inflation never collapse at the ZLB: they are bounded below by the ZLB steady state. Second, when the ZLB episode is long enough, the economy necessarily experiences a slump, irrespective of the duration of forward guidance. Third, government-spending multipliers are always finite, irrespective of the duration of the ZLB episode.

Apart from its anomalies, the standard New Keynesian model has several other intriguing properties at the ZLB---some labeled ``paradoxes'' because they defy usual economic logic \cp{E10,W12,EK11}. Our model shares these properties. First, the paradox of thrift holds: when households desire to save more than their neighbors, the economy contracts and they end up saving the same amount as the neighbors. The paradox of toil also holds: when households desire to work more, the economy contracts and they end up working less. The paradox of flexibility is present too: the economy contracts when prices become more flexible. Last, the government-spending multiplier is above one, so government spending stimulates private consumption.

\section{Justification for Wealth in the Utility Function} 

Before delving into the model, we justify our assumption of wealth in the utility function.

The standard model assumes that people save to smooth consumption over time, but it has long been recognized that people seem to enjoy accumulating wealth irrespective of future consumption. Describing the European upper class of the early 20th century, \ct[chap.~2]{K19} noted that  ``The duty of saving became nine-tenths of virtue and the growth of the cake the object of true religion\ldots. Saving was for old age or for your children; but this was only in theory---the virtue of the cake was that it was never to be consumed, neither by you nor by your children after you.'' Irving Fisher added that ``A man may include in the benefits of his wealth \ldots the social standing he thinks it gives him, or political power and influence, or the mere miserly sense of possession, or the satisfaction in the mere process of further accumulation'' \cp[p.~17]{F30}. Fisher's perspective is interesting since he developed the theory of saving based on consumption smoothing.

Neuroscientific evidence confirms that wealth itself provides utility, independently of the consumption it can buy. \ct[p.~32]{CLP05} note that ``brain-scans conducted while people win or lose money suggest that money activates similar reward areas as do other `primary reinforcers' like food and drugs, which implies that money confers direct utility, rather than simply being valued only for what it can buy.''

Among all the reasons why people may value wealth, we focus on social status: we postulate that people enjoy wealth because it provides social status. We therefore introduce relative (not absolute) wealth into the utility function.\footnote{\ct{CMP92,CMP95} develop models in which relative wealth does not directly confer utility but has other attributes such that people behave as if wealth entered their utility function. In one such model, wealthier individuals have higher social rankings, which allows them to marry wealthier partners and enjoy higher utility.} The assumption is convenient: in equilibrium everybody is the same, so relative wealth is zero. And the assumption seems plausible. Adam Smith, Ricardo, John Rae, J.S. Mill, Marshall, Veblen, and Frank Knight all believed that people accumulate wealth to attain high social status \cp{S81}. More recently, a broad literature has documented that people seek to achieve high social status, and that accumulating wealth is a prevalent pathway to do so \cp{WF98,HF11,F10,AHH15,CT13,R14,MKC17}.\footnote{The wealth-in-the-utility assumption has been found useful in models of long-run growth \cp{K68,K92,Z94,CJ97,FS98}, risk attitudes \cp{R92,C04}, asset pricing \cp{BC96,GZ02,Ka08,MOS18}, life-cycle consumption \cp{Z95,C00,Fra09,S19}, social stratification \cp{LS04}, international macroeconomics \cp{Fi05,FH05}, financial crises \cp{KRW15}, and optimal taxation \cp{SS16}. Such usefulness lends further support to the assumption.}

\section{New Keynesian Model with Wealth in the Utility Function}

We extend the New Keynesian model by assuming that households derive utility not only from consumption and leisure but also from relative wealth. To simplify derivations and be able to represent the equilibrium with phase diagrams, we use an alternative formulation of the New Keynesian model, inspired by \ct{BeSU01} and \ct{W12}. Our formulation features continuous time instead of discrete time; self-employed households instead of firms and households; and \ct{R82} pricing instead of \ct{C83} pricing.
 
\subsection{Assumptions}

The economy is composed of a measure 1 of self-employed households. Each household~$j\in[0,1]$ produces $y_{j}(t)$ units of a good $j$ at time $t$, sold to other households at a price $p_{j}(t)$. The household's production function is $y_{j}(t) = a h_{j}(t)$, where $a>0$ represents the level of technology, and $h_{j}(t)$ is hours of work. Working causes a disutility $\k h_{j}(t)$, where $\k>0$ is the marginal disutility of labor.

The goods produced by households are imperfect substitutes for one another, so each household exercises some monopoly power. Moreover, households face a quadratic cost when they change their price: changing a price at a rate $\pi_{j}(t) = \dot{p}_{j}(t)/p_{j}(t)$ causes a disutility $\g \pi_{j}(t)^2/2$. The parameter $\g > 0$ governs the cost to change prices and thus price rigidity.

Each household consumes goods produced by other households. Household~$j$ buys quantities $c_{jk}(t)$ of the goods $k\in[0,1]$. These quantities are aggregated into a consumption index 
\begin{equation*}
c_{j}(t) = \bs{\int_{0}^{1} c_{jk}(t)^{(\e-1)/\e}\,dk}^{\e/(\e-1)},
\end{equation*}
where $\e>1$ is the elasticity of substitution between goods. The consumption index yields utility $\ln(c_{j}(t))$. Given the consumption index, the relevant price index is 
\begin{equation*}
p(t) = \bs{\int_{0}^1 p_{j}(t)^{1-\e}\,di}^{1/(1-\e)}.
\end{equation*}
When households optimally allocate their consumption expenditure across goods, $p(t)$ is the price of one unit of consumption index. The inflation rate is defined as $\pi(t)= \dot{p}(t)/p(t)$.

Households save using government bonds. Since we postulate that people derive utility from their relative real wealth, and since bonds are the only store of wealth, holding bonds directly provides utility. Formally, holding a nominal quantity of bonds $b_{j}(t)$ yields utility 
\begin{equation*}
u\of{\frac{b_{j}(t)-b(t)}{p(t)}}. 
\end{equation*}
The function $u: \R \to \R$ is increasing and concave, $b(t) = \int_0^1 b_{k}(t)\,dk$ is average nominal wealth, and $[b_{j}(t)-b(t)]/p(t)$ is household~$j$'s relative real wealth.

Bonds earn a nominal interest rate $i^h(t) = i(t) + \s$, where $i(t)\geq 0$ is the nominal interest rate set by the central bank, and $\s\geq 0$ is a spread between the monetary-policy rate ($i(t)$) and the rate used by households for savings decisions ($i^h(t)$). The spread $\s$ captures the efficiency of financial intermediation \cp{W11}; the spread is large when financial intermediation is severely disrupted, as during the Great Depression and Great Recession. The law of motion of household~$j$ bond holdings is
\begin{equation*}
\dot{b}_{j}(t)= i^h(t) b_{j}(t) + p_{j}(t)  y_{j}(t) - \int_0^1 p_{k}(t) c_{jk}(t)\,dk - \tau(t).
\end{equation*}
The term $i^h(t) b_{j}(t)$ is interest income; $p_{j}(t)  y_{j}(t)$ is labor income; $\int_0^1 p_{k}(t) c_{jk}(t)\,dk$ is consumption expenditure; and $\tau(t)$ is a lump-sum tax (used among other things to service government debt). 

Lastly, the problem of household~$j$ is to choose time paths for $y_{j}(t)$, $p_{j}(t)$, $h_{j}(t)$, $\pi_{j}(t)$, $c_{jk}(t)$ for all $k\in[0,1]$, and $b_{j}(t)$ to maximize the discounted sum of instantaneous utilities
\begin{equation*}
\int_{0}^{\infty}e^{-\d t} \bs{\ln(c_{j}(t))+u\of{\frac{b_{j}(t)-b(t)}{p(t)}}- \k h_{j}(t)-\frac{\g}{2} \pi_{j}(t)^2}dt,
\end{equation*}
where $\d>0$ is the time discount rate. The household faces four constraints: production function; law of motion of good $j$'s price, $\dot{p}_{j}(t) = \pi_{j}(t) p_{j}(t)$; law of motion of bond holdings; and demand for good~$j$ coming from other households' maximization, 
\begin{equation*}
y_{j}(t) = \bs{\frac{p_{j}(t)}{p(t)}}^{-\e} c(t),
\end{equation*}
where $c(t) = \int_0^1 c_{k}(t)\,dk$ is aggregate consumption. The household also faces a borrowing constraint preventing Ponzi schemes. The household takes as given aggregate variables, initial wealth $b_{j}(0)$, and initial price $p_{j}(0)$. All households face the same initial conditions, so they will behave the same. 

\subsection{Euler Equation and Phillips Curve}

The equilibrium is described by a system of two differential equations: an Euler equation and a Phillips curve. The Euler-Phillips system governs the dynamics of output $y(t)$ and inflation $\pi(t)$. Here we present the system; formal and heuristic derivations are in appendices~A and B; a discrete-time version is in appendix~C.

The Phillips curve arises from households' optimal pricing decisions:
\begin{equation}
\dot{\pi}(t) = \d \pi(t) -  \frac{\e\k}{\g a}  \bs{y(t) - y^n},
\label{e:phillips}\end{equation}
where
\begin{equation}
y^n = \frac{\e-1}{\e}\cdot \frac{a}{\k}.
\label{e:yn}\end{equation}
The Phillips curve is not modified by wealth in the utility function.

The steady-state Phillips curve, obtained by setting $\dot{\pi}=0$ in \eqref{e:phillips}, describes inflation as a linearly increasing function of output:
\begin{equation}
\pi =  \frac{\e\k}{\d\g a}  \bp{y - y^n}.
\label{e:phillipsss}\end{equation}
We see that $y^n$ is the natural level of output: the level at which producers keep their prices constant.

The Euler equation arises from households' optimal consumption-savings decisions:
\begin{equation}
\frac{\dot{y}(t)}{y(t)}=r(t)- r^n + u'(0) \bs{y(t) - y^n},
\label{e:euler}\end{equation}
where $r(t) = i(t) - \pi(t)$ is the real monetary-policy rate and
\begin{equation}
r^n = \d - \s - u'(0) y^n.
\label{e:rn}\end{equation}

The marginal utility of wealth, $u'(0)$, enters the Euler equation, so unlike the Phillips curve, the Euler equation is modified by the wealth-in-the-utility assumption. To understand why consumption-savings choices are affected by the assumption, we rewrite the Euler equation as
\begin{equation}
\frac{\dot{y}(t)}{y(t)}=r^h(t)- \d + u'(0) y(t),
\label{e:euleri}\end{equation}
where $r^h(t) = r(t) + \s$ is the real interest rate on bonds. In the standard equation, consumption-savings choices are governed by the financial returns on wealth, given by $r^h(t)$, and the cost of delaying consumption, given by $\d$. Here, people also enjoy holding wealth, so a new term appears to capture the hedonic returns on wealth: the marginal rate of substitution between wealth and consumption, $u'(0) y(t)$. In the marginal rate of substitution, the marginal utility of wealth is $u'(0)$ because in equilibrium all households hold the same wealth so relative wealth is zero; the marginal utility of consumption is $1/y(t)$ because consumption utility is log. Thus the wealth-in-the-utility assumption operates by transforming the rate of return on wealth from $r^h(t)$ to $r^h(t)+u'(0)y(t)$. 
 
Because consumption-savings choices depend not only on interest rates but also on the marginal rate of substitution between wealth and consumption, future interest rates have less impact on today's consumption than in the standard model: the Euler equation is discounted. In fact, the discrete-time version of Euler equation \eqref{e:euler} features discounting exactly as in \ct{MNS17} (see appendix~C).

The steady-state Euler equation is obtained by setting $\dot{y}=0$ in \eqref{e:euler}:
\begin{equation}
u'(0) (y-y^n) = r^n - r.
\label{e:eulerss}\end{equation}
The equation describes output as a linearly decreasing function of the real monetary-policy rate---as in the old-fashioned, Keynesian IS curve.  We see that $r^n$ is the natural rate of interest: the real monetary-policy rate at which households consume a quantity $y^n$. 

The steady-state Euler equation is deeply affected by the wealth-in-the-utility assumption. To understand why, we rewrite \eqref{e:eulerss} as
\begin{equation}
r^h+ u'(0) y = \d.
\label{e:eulerssi}\end{equation}
The standard steady-state Euler equation boils down to $r^h= \d$. It imposes that the financial rate of return on wealth equals the time discount rate---otherwise households would not keep their consumption constant. With wealth in the utility function, the returns on wealth are not only financial but also hedonic. The total rate of return becomes $r^h+ u'(0) y$, where the hedonic returns are measured by $u'(0) y$. The steady-state Euler equation imposes that the total rate of return on wealth equals the time discount rate, so it now involves output $y$. When the real interest rate $r^h$ is higher, people have a financial incentive to save more and postpone consumption. They keep consumption constant only if the hedonic returns on wealth fall enough to offset the increase in financial returns: this requires output to decline. As a result, with wealth in the utility function, the steady-state Euler equation describes output as a decreasing function of the real interest rate---as in the traditional IS curve, but through a different mechanism.

The wealth-in-the-utility assumption adds one parameter to the equilibrium conditions: $u'(0)$. Accordingly, we compare two submodels:

\begin{definition} The New Keynesian (NK) model has zero marginal utility of wealth: $u'(0)=0$. The wealth-in-the-utility New Keynesian (WUNK) model has sufficient marginal utility of wealth:
\begin{equation}
u'(0)> \frac{\e\k}{\d\g a}.
\label{e:wunk}\end{equation}\end{definition}

The NK model is the standard model; the WUNK model is the extension proposed in this paper. When prices are fixed ($\g\to \infty$), condition~\eqref{e:wunk} becomes $u'(0)>0$; when prices are perfectly flexible ($\g=0$), condition~\eqref{e:wunk} becomes $u'(0)>\infty$. Hence, at the fixed-price limit, the WUNK model only requires an infinitesimal marginal utility of wealth; at the flexible-price limit, the WUNK model is not well-defined. In the WUNK model we also impose $\d>\sqrt{(\e-1)/\g}$ in order to accommodate positive natural rates of interest.\footnote{Indeed, using \eqref{e:yn} and \eqref{e:wunk}, we see that in the WUNK model
\begin{equation*}
\frac{u'(0) y^n}{\d} > \frac{1}{\d^2}\cdot\frac{\e-1}{\g}.
\end{equation*}
This implies that the natural rate of interest, $r^n = \d \bs{1- u'(0)y^n/\d}$, is bounded above:
\begin{equation*}
r^n < \d  \bs{1 - \frac{1}{\d^2}\cdot\frac{\e-1}{\g}}.
\end{equation*}
For the WUNK model to accommodate positive natural rates of interest, the upper bound on the natural rate must be positive, which requires $\d>\sqrt{(\e - 1)/\g}$.}

\subsection{Natural Rate of Interest and Monetary Policy}

The central bank aims to maintain the economy at the natural steady state, where inflation is at zero and output at its natural level. 

In normal times, the natural rate of interest is positive, and the central bank is able to maintain the economy at the natural steady state using the simple policy rule $i(\pi(t))= r^n + \f \pi(t)$. The corresponding real policy rate is $r(\pi(t)) = r^n + \bp{\f-1} \pi(t)$. The parameter $\f\geq 0$ governs the response of interest rates to inflation: monetary policy is active when $\f>1$ and passive when $\f<1$.

When the natural rate of interest is negative, however, the natural steady state cannot be achieved---because this would require the central bank to set a negative nominal policy rate, which would violate the ZLB. In that case, the central bank moves to the ZLB: $i(t)=0$, so $r(t) = -\pi(t)$.

What could cause the natural rate of interest to be negative? A first possibility is a banking crisis, which disrupts financial intermediation and raises the interest-rate spread \cp{W11,E11}. The natural rate of interest turns negative when the spread is large enough: $\s > \d - u'(0)y^n$.  Another possibility in the WUNK model is drop in consumer sentiment, which leads households to favor saving over consumption, and can be parameterized by an increase in the marginal utility of wealth. The natural rate of interest turns negative when the marginal utility is large enough: $u'(0)>(\d-\s)/y^n$.

\subsection{Properties of the Euler-Phillips System}

We now establish the properties of the Euler-Phillips systems in the NK and WUNK models by constructing their phase diagrams.\footnote{The properties are rederived using an algebraic approach in appendix~D.} The diagrams are displayed in figure~\ref{f:phase}.

\begin{figure}[p]
\subcaptionbox{NK model: normal times, active monetary policy}{\includegraphics[scale=0.21,page=1]{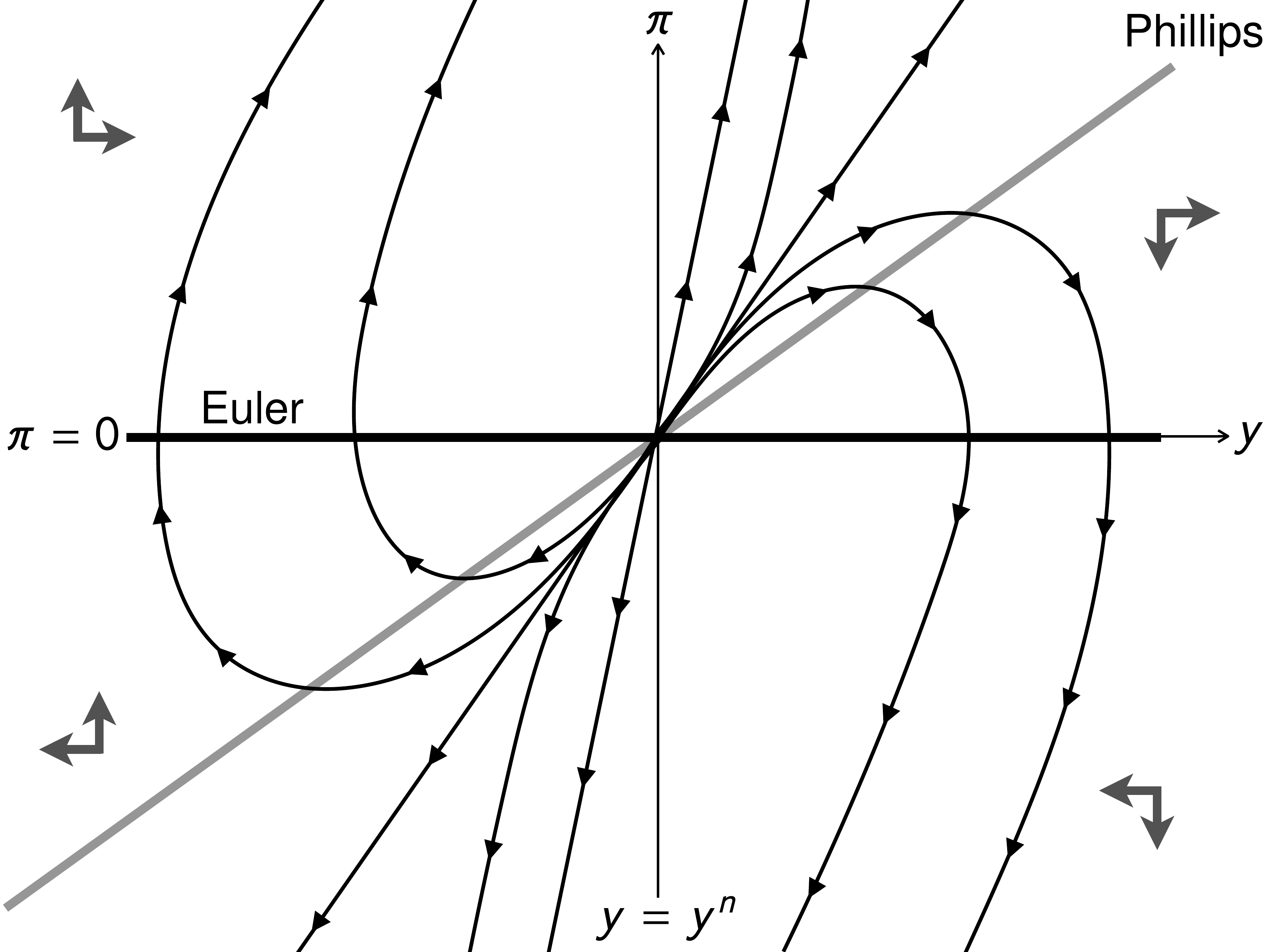}}\hfill
\subcaptionbox{WUNK model: normal times, active monetary policy}{\includegraphics[scale=0.21,page=3]{\pdf}}\vspace{0.5cm}
\subcaptionbox{NK model: ZLB}{\includegraphics[scale=0.21,page=2]{\pdf}}\hfill
\subcaptionbox{WUNK model: ZLB}{\includegraphics[scale=0.21,page=4]{\pdf}}
\caption{Phase Diagrams of the Euler-Phillips System in the NK and WUNK Models}
\note{The figure displays phase diagrams for the dynamical system generated by the Euler equation \eqref{e:euler} and Phillips curve \eqref{e:phillips}: $y$ is output; $\pi$ is inflation; $y^n$ is the natural level of output; the Euler line is the locus $\dot{y}=0$; the Phillips line is the locus $\dot{\pi}=0$; the trajectories are solutions to the Euler-Phillips system linearized around its steady state, plotted for $t$ going from $-\infty$ to $+\infty$. The four panels contrast various cases. The NK model is the standard New Keynesian model. The WUNK model is the same model, except that the marginal utility of wealth is not zero but is sufficiently large to satisfy \eqref{e:wunk}. In normal times, the natural rate of interest $r^n$ is positive, and the monetary-policy rate is given by $i=r^n+\f\pi$; when monetary policy is active, $\f>1$. At the ZLB, the natural rate of interest is negative, and the monetary-policy rate is zero. The figure shows that in the NK model, the Euler-Phillips system is a source in normal times with active monetary policy (panel~A); but the system is a saddle at the ZLB (panel~C). In the WUNK model, by contrast, the Euler-Phillips system is a source both in normal times and at the ZLB (panels~B and D). (Panels~A and B display a nodal source, but the system could also be a spiral source, depending on the value of $\f$; in panel~D the system is always a nodal source.)}
\label{f:phase}\end{figure}

We begin with the Phillips curve, which gives $\dot{\pi}$. First, we plot the locus $\dot{\pi}=0$, which we label ``Phillips.'' The locus is given by the steady-state Phillips curve \eqref{e:phillipsss}: it is linear, upward sloping, and goes through the point $[y=y^n,\pi=0]$. Second, we plot the arrows giving the directions of the trajectories solving the Euler-Phillips system. The sign of $\dot{\pi}$ is given by the Phillips curve \eqref{e:phillips}: any point above the Phillips line (where $\dot{\pi}=0$) has $\dot{\pi}>0$, and any point below the line has $\dot{\pi}<0$. So inflation is rising above the Phillips line and falling below it.

We next turn to the Euler equation, which gives $\dot{y}$. Whereas the Phillips curve is the same in the NK and WUNK models, and in normal times and at the ZLB, the Euler equation is different in each case. We therefore proceed case by case.

We start with the NK model in normal times and with active monetary policy (panel~A). The Euler equation \eqref{e:euler} becomes
\begin{equation*}
\frac{\dot{y}}{y}=(\f-1)\pi,
\end{equation*}
with $\f>1$. The locus $\dot{y}=0$, labeled ``Euler,'' is simply the horizontal line $\pi=0$. Since the Phillips and Euler lines only intersect at the point $[y=y^n,\pi=0]$, we conclude that the Euler-Phillips system admits a unique steady state with zero inflation and natural output. Next we examine the sign of $\dot{y}$. As $\f>1$, any point above the Euler line has $\dot{y}>0$, and any point below the line has $\dot{y}<0$. Since all the trajectories solving the Euler-Phillips system move away from the steady state in the four quadrants delimited by the Phillips and Euler lines, we conclude that the Euler-Phillips system is a source.

We then consider the WUNK model in normal times with active monetary policy (panel~B). The Euler equation \eqref{e:euler} becomes
\begin{equation*}
\frac{\dot{y}}{y}=(\f-1)\pi + u'(0)\bp{y-y^n},
\end{equation*}
with $\f>1$. We first use the Euler equation to compute the Euler line (locus $\dot{y}=0$):
\begin{equation*}
\pi = -\frac{u'(0)}{\f-1}(y-y^n).
\end{equation*}
The Euler line is linear, downward sloping (as $\f>1$), and goes through the point $[y=y^n,\pi=0]$. Since the Phillips and Euler lines only intersect at the point $[y=y^n,\pi=0]$, we conclude that the Euler-Phillips system admits a unique steady state, with zero inflation and output at its natural level. Next we use the Euler equation to determine the sign of $\dot{y}$. As $\f>1$, any point above the Euler line has $\dot{y}>0$, and any point below it has $\dot{y}<0$. Hence, the solution trajectories move away from the steady state in all four quadrants of the phase diagram; we conclude that the Euler-Phillips system is a source. In normal times with active monetary policy, the Euler-Phillips system therefore behaves similarly in the NK and WUNK models.

We next turn to the NK model at the ZLB (panel~C). The Euler equation \eqref{e:euler} becomes
\begin{equation*}
\frac{\dot{y}}{y}=-\pi-r^n.
\end{equation*}
Thus the Euler line (locus $\dot{y}=0$) shifts up from $\pi=0$ in normal times to $\pi=-r^n>0$ at the ZLB.  We infer that the Euler-Phillips system admits a unique steady state, where inflation is positive and output is above its natural level. Furthermore, any point above the Euler line has $\dot{y}<0$, and any point below it has $\dot{y}>0$. As a result, the solution trajectories move toward the steady state in the southwest and northeast quadrants of the phase diagram, whereas they move away from it in the southeast and northwest quadrants. We infer that the Euler-Phillips system is a saddle. 

We finally move to the WUNK model at the ZLB (panel~D). The Euler equation \eqref{e:euler} becomes
\begin{equation*}
\frac{\dot{y}}{y}=-\pi-r^n + u'(0)\bp{y-y^n}.
\end{equation*}
First, this differential equation implies that the Euler line (locus $\dot{y}=0$) is given by
\begin{equation}
\pi=-r^n+u'(0)(y-y^n).
\label{e:eulerzlb}\end{equation}
So the Euler line is linear, upward sloping, and goes through the point $[y=y^n+r^n/u'(0),\pi=0]$. The Euler line has become upward sloping because the real monetary-policy rate, which was increasing with inflation when monetary policy was active, has become decreasing with inflation at the ZLB ($r=-\pi$). Since $r^n\leq 0$, the Euler line has shifted inward of the point $[y=y^n,\pi=0]$, explaining why the central bank is unable to achieve the natural steady state at the ZLB. And since the slope of the Euler line is $u'(0)$ while that of the Phillips line is $\e\k/(\d\g a)$, the WUNK condition \eqref{e:wunk} ensures that the Euler line is steeper than the Phillips line at the ZLB. From the Euler and Phillips lines, we infer that the Euler-Phillips system admits a unique steady state, in which inflation is negative and output is below its natural level.\footnote{We also check that the intersection of the Euler and Phillips lines has positive output (appendix~D).}

Second, the differential equation shows that any point above the Euler line has $\dot{y}<0$, and any point below it has $\dot{y}>0$. Hence, in all four quadrants of the phase diagram, the trajectories move away from the steady state. We conclude that the Euler-Phillips system is a source. At the ZLB, the Euler-Phillips system therefore behaves very differently in the NK and WUNK models.

With passive monetary policy in normal times, the phase diagrams of the Euler-Phillips system would be similar to the ZLB phase diagrams---except that the Euler and Phillips lines would intersect at $[y=y^n,\pi=0]$. In particular, the Euler-Phillips system would be a saddle in the NK model and a source in the WUNK model.

For completeness, we also plot sample solutions to the Euler-Phillips system. The trajectories are obtained by linearizing the Euler-Phillips system at its steady state.\footnote{Technically the trajectories only approximate the exact solutions; but the approximation is accurate in the neighborhood of the steady state.} When the system is a source, there are two unstable lines (trajectories that move away from the steady state in a straight line). At $t\to -\infty$, all other trajectories are in the vicinity of the steady state and move away tangentially to one of the unstable lines. At $t\to+\infty$, the trajectories move to infinity parallel to the other unstable line. When the system is a saddle, there is one unstable line and one stable line (trajectory that goes to the steady state in a straight line). All other trajectories come from infinity parallel to the stable line when $t\to -\infty$, and move to infinity parallel to the unstable line when $t\to +\infty$.

The next propositions summarize the results:

\begin{proposition}\label{p:normal} Consider the Euler-Phillips system in normal times. The system admits a unique steady state, where output is at its natural level, inflation is zero, and the ZLB is not binding. In the NK model, the system is a source when monetary policy is active but a saddle when monetary policy is passive. In the WUNK model, the system is a source whether monetary policy is active or passive.\end{proposition}

\begin{proposition}\label{p:zlb} Consider the Euler-Phillips system at the ZLB. In the NK model, the system admits a unique steady state, where output is above its natural level and inflation is positive; furthermore, the system is a saddle. In the WUNK model, the system admits a unique steady state, where output is below its natural level and inflation is negative; furthermore, the system is a source.\end{proposition}

The propositions give the key difference between the NK and WUNK models: at the ZLB, the Euler-Phillips system remains a source in the WUNK model, whereas it becomes a saddle in the NK model. This difference will explain why the WUNK model does not suffer from the anomalies plaguing the NK model at the ZLB. The phase diagrams also illustrate the origin of the WUNK condition \eqref{e:wunk}. In the WUNK model, the Euler-Phillips system remains a source at the ZLB as long as the Euler line is steeper than the Phillips line (figure~\ref{f:phase}, panel~D). The Euler line's slope at the ZLB is the marginal utility of wealth, so that marginal utility is required to be above a certain level---which is given by~\eqref{e:wunk}.

The propositions have implications for equilibrium determinacy. When the Euler-Phillips system is a source, the equilibrium is determinate: the only equilibrium trajectory in the vicinity of the steady state is to jump to the steady state and stay there; if the economy jumped somewhere else, output or inflation would diverge following a trajectory similar to those plotted in panels~A, B, and D of figure~\ref{f:phase}. When the system is a saddle, the equilibrium is indeterminate: any trajectory jumping somewhere on the saddle path and converging to the steady state is an equilibrium (figure~\ref{f:phase}, panel~C). Hence, in the NK model, the equilibrium is determinate when monetary policy is active but indeterminate when monetary policy is passive and at the ZLB. In the WUNK model, the equilibrium is always determinate, even when monetary policy is passive and at the ZLB.

Accordingly, in the NK model, the Taylor principle holds: the central bank must adhere to an active monetary policy to avoid indeterminacy. From now on, we therefore assume that the central bank in the NK model follows an active policy whenever it can ($\f>1$ whenever $r^n>0$). In the WUNK model, by contrast, indeterminacy is never a risk, so the central bank does not need to worry about how strongly its policy rate responds to inflation. The central bank could even follow an interest-rate peg without creating indeterminacy.

The results that pertain to the NK model in propositions~\ref{p:normal} and \ref{p:zlb} are well-known \eg{W01}. The results that pertain to the WUNK model are close to those obtained by \ct[proposition~3.1]{G16}, although he does not use our phase-diagram representation. Gabaix finds that when bounded rationality is strong enough, the Euler-Phillips system is a source even at the ZLB. He also finds that when prices are more flexible, more bounded rationality is required to maintain the source property. The same is true here: when the marginal utility of wealth is high enough, such that \eqref{e:wunk} holds, the Euler-Phillips system is a source even at the ZLB; and when the price-adjustment cost $\g$ is lower, \eqref{e:wunk} imposes a higher threshold on the marginal utility of wealth. Our phase diagrams illustrate the logic behind these results. The Euler-Phillips system remains a source at the ZLB as long as the Euler line is steeper than the Phillips line (figure~\ref{f:phase}, panel~D). As the slope of the Euler line is determined by bounded rationality in the Gabaix model and by marginal utility of wealth in our model, these need to be large enough. When prices are more flexible, the Phillips line steepens, so the Euler line's required steepness increases: bounded rationality or marginal utility of wealth need to be larger.

\section{Description and Resolution of the New Keynesian Anomalies}\label{s:anomalies}

We now describe the anomalous predictions of the NK model at the ZLB: implausibly large drop in output and inflation; and implausibly strong effects of forward guidance and government spending. We then show that these anomalies are absent from the WUNK model.

\begin{table}[p]
\caption{ZLB Scenarios}
\normalsize\begin{tabular*}{\textwidth}[]{p{3.3cm}@{\extracolsep\fill}cccc}
\toprule
& Timeline & Natural rate & Monetary  & Government \\
&  & of interest &  policy & spending\\
\midrule
\multicolumn{5}{l}{A. ZLB episode}\\
\midrule
ZLB: & $t\in (0,T) $ & $r^n<0$ & $i=0$  & -- \\
Normal times: & $t>T$ &  $r^n>0$ & $i= r^n + \f \pi$  & --   \\
\midrule
\multicolumn{5}{l}{B. ZLB episode with forward guidance}\\
\midrule
ZLB: & $t\in (0,T)$ & $r^n<0$ & $i=0$  & -- \\
Forward guidance: & $t\in (T,T+\D)$ & $r^n>0$ & $i=0$  & --\\
Normal times: & $t>T+\D$ & $r^n>0$ & $i= r^n + \f \pi$  & -- \\
\midrule
\multicolumn{5}{l}{C. ZLB episode with government spending}\\
\midrule
ZLB: & $t\in (0,T) $ & $r^n<0$ & $i=0$  & $g>0$  \\
Normal times: & $t>T$ & $r^n>0$ & $i= r^n + \f \pi$  & $g=0$   \\
\bottomrule
\end{tabular*}
\note{This table describes the three scenarios analyzed in section~4: the ZLB episode, in section 4.1; the ZLB episode with forward guidance, in section 4.2; and the ZLB episode with government spending, in section 4.3. The parameter $T>0$ gives the duration of the ZLB episode; the parameter $\D>0$ gives the duration of forward guidance. We assume that monetary policy is active ($\f>1$) in normal times in the NK model; this assumption is required to ensure equilibrium determinacy (Taylor principle). In the WUNK model, monetary policy can be active or passive in normal times.}
\label{t:scenarios}\end{table}

\subsection{Drop in Output and Inflation}

\begin{figure}[p]
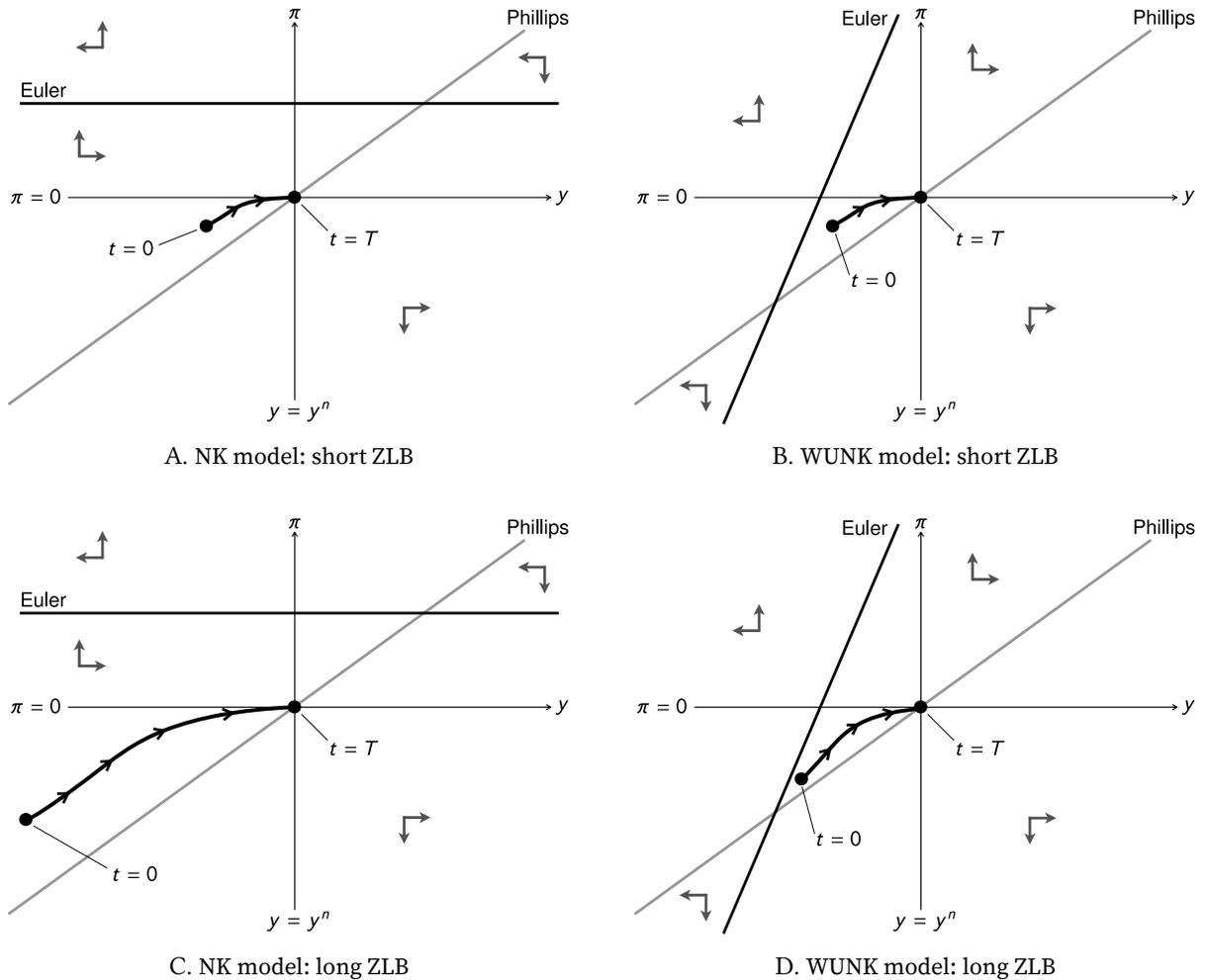

\subcaptionbox{NK model: short ZLB}{\includegraphics[scale=0.21,page=5]{\pdf}}\hfill
\subcaptionbox{WUNK model: short ZLB}{\includegraphics[scale=0.21,page=7]{\pdf}}\vspace{0.5cm}
\subcaptionbox{NK model: long ZLB}{\includegraphics[scale=0.21,page=6]{\pdf}}\hfill
\subcaptionbox{WUNK model: long ZLB}{\includegraphics[scale=0.21,page=8]{\pdf}}
\caption{ZLB Episodes in the NK and WUNK Models}
\note{The figure describes various ZLB episodes. The timeline of a ZLB episode is presented in table~\ref{t:scenarios}, panel~A. Panel A displays the phase diagram of the NK model's Euler-Phillips system at the ZLB; it comes from figure~\ref{f:phase}, panel~C. Panel B displays the phase diagram of the WUNK model's Euler-Phillips system at the ZLB; it comes from figure~\ref{f:phase}, panel~D. Panels C and D are the same as panels~A and B, but with a longer-lasting ZLB (larger $T$). The equilibrium trajectories are the unique trajectories reaching the natural steady state (where $\pi=0$ and $y=y^n$) at time~$T$. The figure shows that the economy slumps during the ZLB: inflation is negative and output is below its natural level (panels~A and B). In the NK model, the initial slump becomes unboundedly severe as the ZLB lasts longer (panel~C). In the WUNK model, there is no such collapse: output and inflation are bounded below by the ZLB steady state (panel~D).}
\label{f:collapse}\end{figure}

We consider a temporary ZLB episode, as in \ct{W12}. Between times $0$ and $T>0$, the natural rate of interest is negative. In response, the central bank maintains its policy rate at zero. After time $T$, the natural rate is positive again, and monetary policy returns to normal. This scenario is summarized in table~\ref{t:scenarios}, panel~A. We analyze the ZLB episode using the phase diagrams in figure~\ref{f:collapse}.

We start with the NK model. We analyze the ZLB episode by going backward in time. After time $T$, monetary policy maintains the economy at the natural steady state. Since equilibrium trajectories are continuous, the economy also is at the natural steady state at the end of the ZLB, when $t=T$.\footnote{The trajectories are continuous in output and inflation because households have concave preferences over the two arguments. If consumption had an expected discrete jump, for example, households would be able to increase their utility by reducing the size of the discontinuity.} 

We then move back to the ZLB episode, when $t<T$. At time $0$, the economy jumps to the unique position leading to $[y=y^n,\pi=0]$ at time $T$. Hence, inflation and output initially jump down to $\pi(0)<0$ and $y(0)<y^n$, and then recover following the unique trajectory leading to $[y=y^n,\pi=0]$. The ZLB therefore creates a slump, with below-natural output and deflation (panel~A). 

Critically, the economy is always on the same trajectory during the ZLB, irrespective of the ZLB duration $T$. A longer ZLB only forces output and inflation to start from a lower position on the trajectory at time $0$. Thus, as the ZLB lasts longer, initial output and inflation collapse to unboundedly low levels (panel~C). 

Now let us examine the WUNK model. Output and inflation never collapse during the ZLB. Initially inflation and output jump down toward the ZLB steady state, denoted $[y^z,\pi^z]$, so $\pi^z<\pi(0)<0$ and $y^z<y(0)<y^n$. They then recover following the trajectory going through $[y=y^n,\pi=0]$. Consequently the ZLB episode creates a slump (panel~B), which is deeper when the ZLB lasts longer (panel~D). But unlike in the NK model, the slump is bounded below by the ZLB steady state: irrespective of the duration of the ZLB, output and inflation remain above $y^z$ and $\pi^z$, respectively, so they never collapse. Moreover, if the natural rate of interest is negative but close to zero, such that $\pi^z$ is close to zero and $y^z$ to $y^n$, output and inflation will barely deviate from the natural steady state during the ZLB---even if the ZLB lasts a very long time.

The following proposition records these results:\footnote{The result that in the NK model output becomes infinitely negative when the ZLB becomes infinitely long should not be interpreted literally. It is obtained because we omitted the constraint that output must remain positive. The proper interpretation is that output falls much, much below its natural level---in fact it converges to zero.}

\begin{proposition} Consider a ZLB episode between times $0$ and $T$. The economy enters a slump: $y(t)<y^n$ and $\pi(t)<0$ for all $t\in(0,T)$. In the NK model, the slump becomes infinitely severe as the ZLB duration approaches infinity:  $\lim_{T\to \infty} y(0) = \lim_{T\to \infty} \pi(0) = -\infty$. In the WUNK model, in contrast, the slump is bounded below by the ZLB steady state $[y^z,\pi^z]$: $y(t)> y^z$ and $\pi(t)>\pi^z$ for all $t\in(0,T)$. In fact, the slump approaches the ZLB steady state as the ZLB duration approaches infinity: $\lim_{T\to \infty} y(0) = y^z$ and $\lim_{T\to \infty} \pi(0) = \pi^z$.\end{proposition}

In the NK model, output and inflation collapse when the ZLB is long-lasting, which is well-known (\inp[fig.~1]{EW04}; \inp[fig.~1]{E11}; \inp[proposition~1]{W12}). This collapse is difficult to reconcile with real-world observations. The ZLB episode that started in 1995 in Japan lasted for more than twenty years without sustained deflation. The ZLB episode that started in 2009 in the euro area lasted for more than 10 years; it did not yield sustained deflation either. The same is true of the ZLB episode that occurred in the United States between 2008 and 2015.

In the WUNK model, in contrast, inflation and output never collapse. Instead, as the duration of the ZLB increases, the economy converges to the ZLB steady state. That ZLB steady state may not be far from the natural steady state: if the natural rate of interest is only slightly negative, inflation is only slightly below zero and output only slightly below its natural level. \ct[proposition~3.2]{G16} obtains a closely related result: in his model output and inflation also converge to the ZLB steady state when the ZLB is arbitrarily long.

\subsection{Forward Guidance}

We turn to the effects of forward guidance at the ZLB. We consider a three-stage scenario, as in \ct{C17}. Between times $0$ and $T$, there is a ZLB episode. To alleviate the situation, the central bank makes a forward-guidance promise at time $0$: that it will maintain the policy rate at zero for a duration $\D$ once the ZLB is over. After time $T$, the natural rate of interest is positive again. Between times $T$ and $T+\D$, the central bank fulfills its forward-guidance promise and keeps the policy rate at zero. After time $T+\D$, monetary policy returns to normal. This scenario is summarized in table~\ref{t:scenarios}, panel~B.

\begin{figure}[p]
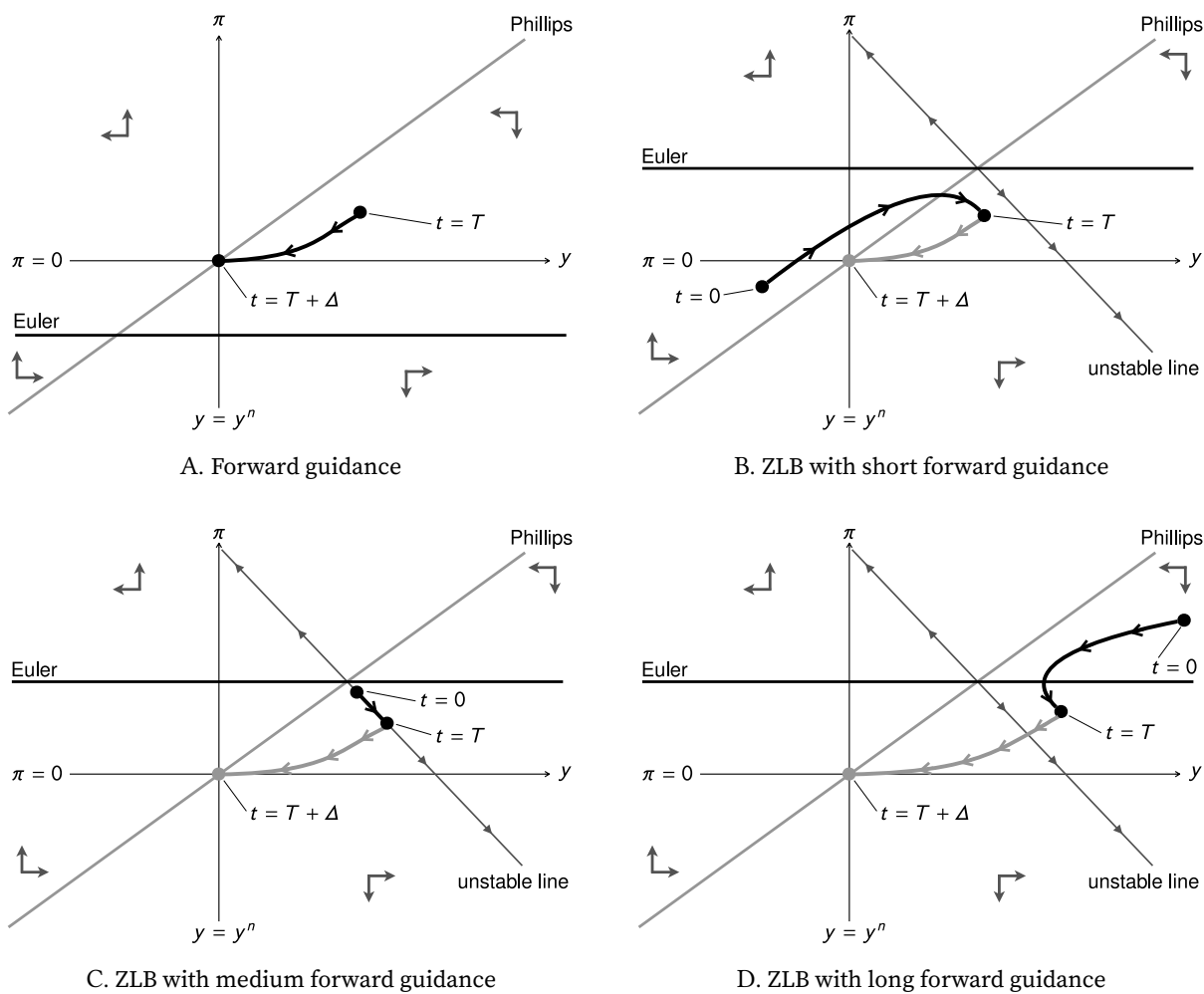

\subcaptionbox{Forward guidance}{\includegraphics[scale=0.21,page=9]{\pdf}}\hfill
\subcaptionbox{ZLB with short forward guidance}{\includegraphics[scale=0.21,page=10]{\pdf}}\vspace{0.5cm}
\subcaptionbox{ZLB with medium forward guidance}{\includegraphics[scale=0.21,page=11]{\pdf}}\hfill
\subcaptionbox{ZLB with long forward guidance}{\includegraphics[scale=0.21,page=12]{\pdf}}
\caption{NK Model: ZLB Episodes with Forward Guidance}
\note{The figure describes various ZLB episodes with forward guidance in the NK model. The timeline of such episode is presented in table~\ref{t:scenarios}, panel~B. Panel A displays the phase diagram of the NK model's Euler-Phillips system during forward guidance; it is similar to the diagram in figure~\ref{f:phase}, panel~C but with $r^n>0$. The equilibrium trajectory during forward guidance is the unique trajectory reaching the natural steady state at time~$T+\D$. Panels~B, C, and D display the phase diagram of the NK model's Euler-Phillips system at the ZLB; they comes from figure~\ref{f:phase}, panel~C. The equilibrium trajectory at the ZLB is the unique trajectory reaching the point determined by forward guidance at time $T$. Panels B, C, and D differ in the underlying duration of forward guidance ($\D$): short in panel~B, medium in panel~C, and long in panel~D. The figure shows that the NK model suffers from an anomaly: when forward guidance lasts sufficiently to bring $[y(T),\pi(T)]$ above the unstable line, any ZLB episode---however long---triggers a boom (panel~D). On the other hand, if forward guidance is short enough to keep $[y(T),\pi(T)]$ below the unstable line, long-enough ZLB episodes are slumps (panel~B). In the knife-edge case where $[y(T),\pi(T)]$ falls just on the unstable line, arbitrarily long ZLB episodes converge to the ZLB steady state (panel~C).}
\label{f:guidancenk}\end{figure}

We analyze the ZLB episode with forward guidance using the phase diagrams in figures~\ref{f:guidancenk} and \ref{f:guidancewunk}. 
The forward-guidance diagrams are based on the ZLB diagrams in figure~\ref{f:phase}. In the NK model (figure~\ref{f:guidancenk}, panel~A), the diagram is the same as in panel~C of figure~\ref{f:phase}, except that the Euler line $\pi=-r^n$ is lower because $r^n>0$ instead of $r^n<0$. In the WUNK model (figure~\ref{f:guidancewunk}, panel~A), the diagram is the same as in panel~D of figure~\ref{f:phase}, except that the Euler line \eqref{e:eulerzlb} is shifted outward because $r^n>0$ instead of $r^n<0$.

We begin with the NK model (figure~\ref{f:guidancenk}). We go backward in time. After time $T+\D$, monetary policy maintains the economy at the natural steady state. Between times $T$ and $T+\D$, the economy is in forward guidance (panel~A). Following the logic of figure~\ref{f:collapse}, we find that at time $T$, inflation is positive and output above its natural level. They subsequently decrease over time, following the unique trajectory leading to the natural steady state at time $T+\D$. Accordingly, the economy booms during forward guidance. Furthermore, as forward guidance lengthens, inflation and output at time $T$ become higher. 

We look next at the ZLB episode, between times $0$ and $T$. Since equilibrium trajectories are continuous, the economy is at the same point at the end of the ZLB and at the beginning of forward guidance. The boom engineered during forward guidance therefore improves the situation at the ZLB. Instead of reaching the natural steady state at time $T$, the economy reaches a point with positive inflation and above-natural output, so at any time before $T$, inflation and output tend to be higher than without forward guidance (panel~B).

Forward guidance can actually have tremendously strong effects in the NK model. For small durations of forward guidance, the position at time $T$ is below the ZLB unstable line. It is therefore connected to trajectories coming from the southwest quadrant of the phase diagram (panel~B). As the ZLB lasts longer, initial output and inflation collapse. When the duration of forward guidance is such that the position at time $T$ is exactly on the unstable line, the position at time $0$ is on the unstable line as well (panel~C). As the ZLB lasts longer, the initial position inches closer to the ZLB steady state. For even longer forward guidance, the position at time $T$ is above the unstable line, so it is connected to trajectories coming from the northeast quadrant (panel~D). Then, as the ZLB lasts longer, initial output and inflation become higher and higher. As a result, if the duration of forward guidance is long enough, a deep slump can be transformed into a roaring boom. Moreover, the forward-guidance duration threshold is independent of the ZLB duration.

\begin{figure}[p]
\subcaptionbox{Forward guidance}{\includegraphics[scale=0.21,page=13]{\pdf}}\hfill
\subcaptionbox{Short ZLB with forward guidance}{\includegraphics[scale=0.21,page=14]{\pdf}}\vspace{0.5cm}
\subcaptionbox{Long ZLB with forward guidance}{\includegraphics[scale=0.21,page=15]{\pdf}}\hfill
\subcaptionbox{Possible trajectories}{\includegraphics[scale=0.21,page=16]{\pdf}}
\caption{WUNK Model: ZLB Episodes with Forward Guidance}
\note{The figure describes various ZLB episodes with forward guidance in the WUNK model. The timeline of such episode is presented in table~\ref{t:scenarios}, panel~B. Panel A displays the phase diagram of the WUNK model's Euler-Phillips system during forward guidance; it is similar to the diagram in figure~\ref{f:phase}, panel~D but with $r^n>0$. The equilibrium trajectory during forward guidance is the unique trajectory reaching the natural steady state at time~$T+\D$. Panel~B displays the phase diagram of the WUNK model's Euler-Phillips system at the ZLB; it comes from figure~\ref{f:phase}, panel~D. The equilibrium trajectory at the ZLB is the unique trajectory reaching the point determined by forward guidance at time $T$. Panel C is the same as panel~B, but with a longer-lasting ZLB (larger $T$). Panel~D is a generic version of panels~A, B, and C, describing any duration of ZLB and forward guidance. The figure shows that the NK model's anomaly disappears in the WUNK model: a long-enough ZLB episode prompts a slump irrespective of the duration of forward guidance (panel~C).}
\label{f:guidancewunk}\end{figure}

In comparison, the power of forward guidance is subdued in the WUNK model (figure~\ref{f:guidancewunk}). Between times $T$ and $T+\D$, forward guidance operates (panel~A). Inflation is positive and output is above its natural level at time $T$. They then decrease over time, following the trajectory leading to the natural steady state at time $T+\D$. The economy booms during forward guidance; but unlike in the NK model, output and inflation are bounded above by the forward-guidance steady state.

Before forward guidance comes the ZLB episode (panels~B and C). Thanks to the boom engineered by forward guidance, the situation is improved at the ZLB: inflation and output tend to be higher than without forward guidance. Yet, unlike in the NK model, output during the ZLB episode is always below its level at time $T$, so forward guidance cannot generate unbounded booms (panel~D). The ZLB cannot generate unbounded slumps either, since output and inflation are bounded below by the ZLB steady state (panel~D). Actually, for any forward-guidance duration, as the ZLB lasts longer, the economy converges to the ZLB steady state at time $0$. The implication is that forward guidance can never prevent a slump when the ZLB lasts long enough.

Based on these dynamics, we identify an anomaly in the NK model, which is resolved in the WUNK model (proof details in appendix~D):

\begin{proposition} Consider a ZLB episode during $(0,T)$ followed by forward guidance during $(T,T+\D)$.
\begin{itemize}
\item In the NK model, there exists a threshold $\D^*$ such that a forward guidance longer than $\D^*$ transforms a ZLB episode of any duration into a boom: let $\D>\D^*$; for any $T$ and for all $t\in(0,T+\D)$, $y(t)>y^n$ and $\pi(t)>0$. In addition, when forward guidance is longer than $\D^*$, a long-enough forward guidance or ZLB episode generates an arbitrarily large boom: for any $T$, $\lim_{\D\to \infty} y(0) = \lim_{\D\to \infty} \pi(0) = +\infty$; and for any $\D>\D^*$, $\lim_{T\to \infty} y(0) = \lim_{T\to \infty} \pi(0) = +\infty$.
\item  In the WUNK model, in contrast, there exists a threshold $T^*$ such that a ZLB episode longer than $T^*$ prompts a slump, irrespective of the duration of forward guidance: let $T>T^*$; for any $\D$, $y(0)<y^n$ and $\pi(0)<0$. Furthermore, the slump approaches the ZLB steady state as the ZLB duration approaches infinity: for any $\D$, $\lim_{T\to \infty} y(0) = y^z$ and $\lim_{T\to \infty} \pi(0) = \pi^z$. In addition, the economy is bounded above by the forward-guidance steady state $[y^f,\pi^f]$: for any $T$ and $\D$, and for all $t\in(0,T+\D)$, $y(t)<y^f$ and $\pi(t)<\pi^f$.
\end{itemize}\end{proposition}

The anomaly identified in the proposition corresponds to the forward-guidance puzzle described by \ct[fig.~1]{CFP15} and \ct[fig.~6]{C17}.\footnote{In the literature the forward-guidance puzzle takes several forms. The common element is that future monetary policy has an implausibly strong effect on current output and inflation.} These papers also find that a long-enough forward guidance transforms a ZLB slump into a boom.

In the WUNK model, this anomalous pattern vanishes. In the New Keynesian models by \ct{G16}, \ct{DL17}, \ct{AD18}, and \ct{B18}, forward guidance also has more subdued effects than in the standard model. Besides, New Keynesian models have been developed with the sole goal of solving the forward-guidance puzzle. Among these, ours belongs to the group that uses discounted Euler equations.\footnote{Other approaches to solve the forward-guidance puzzle include modifying the Phillips curve \cp{CFP15}, combining reflective expectations and temporary equilibrium \cp{GW15}, combining bounded rationality and incomplete markets \cp{FW17}, and introducing an endogenous liquidity premium \cp{BKS18}.} For example, \ct{DGP12} generate discounting from overlapping generations; \ct{MNS16} from heterogeneous agents facing borrowing constraints and cyclical income risk; \ct{AL16} from incomplete information; and \ct{CFJ17} from government bonds in the utility function (which is closely related to our approach).

\subsection{Government Spending}

Last we consider the effects of government spending at the ZLB. We first extend the model by assuming that the government purchases goods from all households, which are aggregated into public consumption $g(t)$. To ensure that government spending affects inflation and private consumption, we also assume that the disutility of labor is convex: household~$j$ incurs disutility $\k^{1+\eta} h_{j}(t)^{1+\eta}/(1+\eta)$ from working, where $\eta> 0$ is the inverse of the Frisch elasticity. Complete extended model, derivations, and results are presented in appendix~E.

In this model, the Euler equation is unchanged, but the Phillips curve is modified because the marginal disutility of labor is not constant, and because households produce goods for the government. The modification of the Phillips curve alters the analysis in three ways. 

First, the steady-state Phillips curve becomes nonlinear, which may introduce additional steady states. We handle this issue as in the literature: we linearize the Euler-Phillips system around the natural steady state without government spending, and concentrate on the dynamics of the linearized system. These dynamics are described by phase diagrams similar to those in the basic model.

Second, the slope of the steady-state Phillips curve is modified, so the WUNK assumption needs to be adjusted. Instead of \eqref{e:phillipsss}, the linearized steady-state Phillips curve is
\begin{equation}
\pi =  - \frac{\e\k}{\d \g a} \bp{\frac{\e-1}{\e}}^{\eta/(1+\eta)}\bs{(1+\eta) (c-c^n) + \eta g}.
\label{e:phillipsgss}\end{equation}
The WUNK assumption guarantees that at the ZLB, the steady-state Euler equation (with slope $u'(0)$) is steeper than the steady-state Phillips curve (now given by \eqref{e:phillipsgss}). Hence, we need to replace assumption \eqref{e:wunk} by
\begin{equation}
u'(0) > (1+\eta) \frac{\e\k}{\d\g a}\bp{\frac{\e-1}{\e}}^{\eta/(1+\eta)}.
\label{e:wunkg}\end{equation}
Naturally, for $\eta=0$, this assumption reduces to \eqref{e:wunk}. 

Third, public consumption enters the Phillips curve, so government spending operates through that curve. Indeed, since $\eta>0$ in \eqref{e:phillipsgss}, government spending shifts the steady-state Phillips curve upward. Intuitively, given private consumption, an increase in government spending raises production and thus marginal costs. Facing higher marginal costs, producers augment inflation.  

\begin{figure}[p]
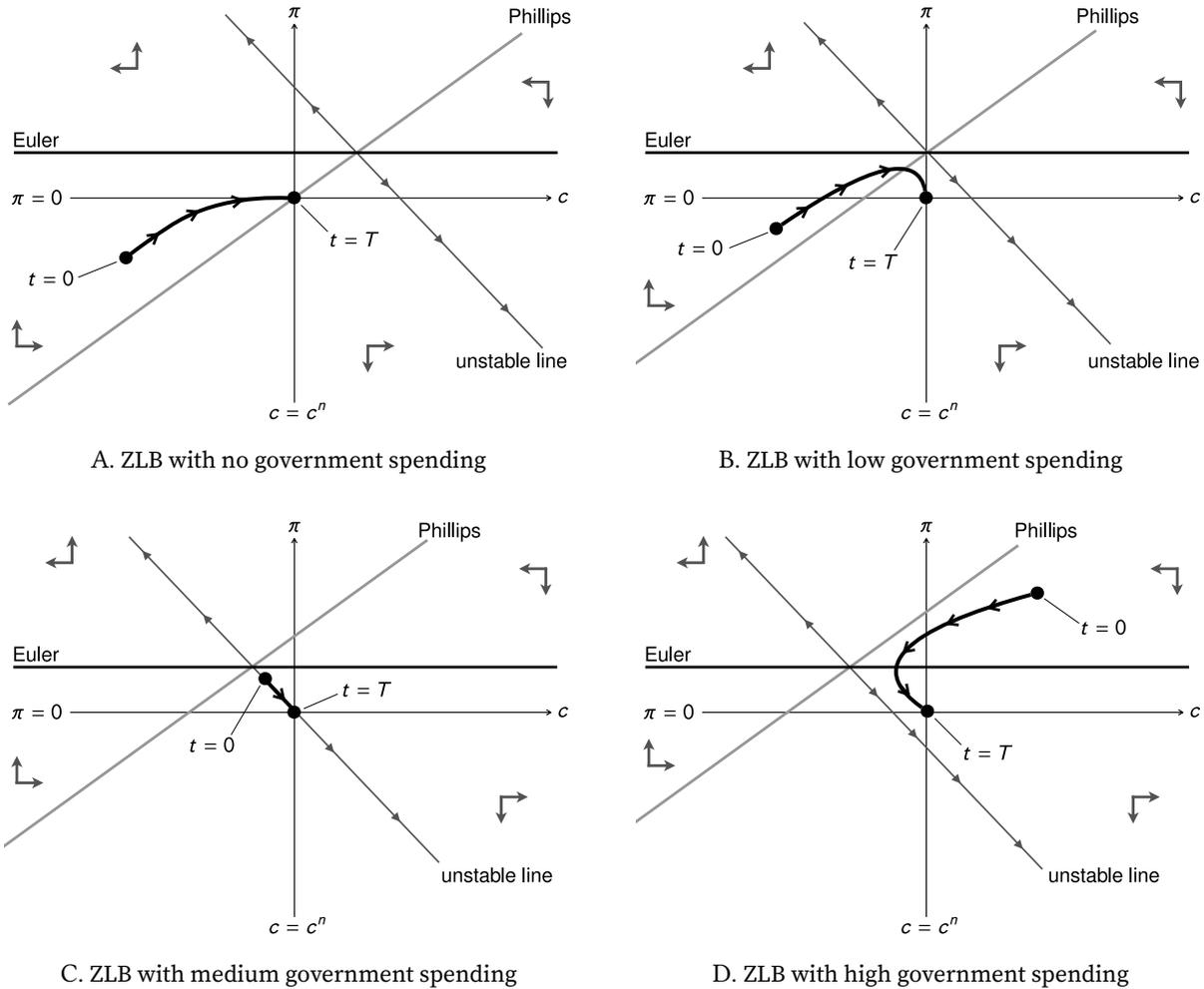

\subcaptionbox{ZLB with no government spending}{\includegraphics[scale=0.21,page=17]{\pdf}}\hfill
\subcaptionbox{ZLB with low government spending}{\includegraphics[scale=0.21,page=18]{\pdf}}\vspace{0.5cm}
\subcaptionbox{ZLB with medium government spending}{\includegraphics[scale=0.21,page=19]{\pdf}}\hfill
\subcaptionbox{ZLB with high government spending}{\includegraphics[scale=0.21,page=20]{\pdf}}
\caption{NK Model: ZLB Episodes with Government Spending}
\note{The figure describes various ZLB episodes with government spending in the NK model. The timeline of such episode is presented in table~\ref{t:scenarios}, panel~C. The panels display the phase diagrams of the linearized Euler-Phillips system for the NK model with government spending and convex disutility of labor at the ZLB: $c$ is private consumption; $\pi$ is inflation; $c^n$ is the natural level of private consumption; the Euler line is the locus $\dot{c}=0$; the Phillips line is the locus $\dot{\pi}=0$. The phase diagrams have the same properties as that in figure~\ref{f:phase}, panel~C, except that the Phillips line shifts upward when government spending increases (see equation \eqref{e:phillipsgss}). The equilibrium trajectory at the ZLB is the unique trajectory reaching the natural steady state at time~$T$. The four panels feature an increasing amount of government spending ($g$), starting from $g=0$ in panel~A. The figure shows that the NK model suffers from an anomaly: when government spending brings down the unstable line from above to below the natural steady state, an arbitrarily long ZLB episode sees an arbitrarily large increase in output, which triggers an unboundedly large boom (from panel~B to panel~D). On the other hand, if government spending is low enough to keep the unstable line above the natural steady state, long-enough ZLB episodes are slumps (panel~B). In the knife-edge case where the natural steady state falls just on the unstable line, arbitrarily long ZLB episodes converge to the ZLB steady state (panel~C).}
\label{f:spendingnk}\end{figure}

We now study a ZLB episode during which the government increases spending in an effort to stimulate the economy, as in \ct{C17}. Between times $0$ and $T$, there is a ZLB episode. To alleviate the situation, the government provides an amount $g>0$ of public consumption. After time $T$, the natural rate of interest is positive again, government spending stops, and monetary policy returns to normal. This scenario is summarized in table~\ref{t:scenarios}, panel~C.

We start with the NK model (figure~\ref{f:spendingnk}).\footnote{There is a small difference with the phase diagrams of the basic model: private consumption $c$ is on the horizontal axis instead of output $y$. But $y=c$ in the basic model (government spending is zero), so the phase diagrams with private consumption on the horizontal axis would be the same as those with output.} We construct the equilibrium path by going backward in time. At time $T$, monetary policy brings the economy to the natural steady state. At the ZLB, government spending helps, but through a different mechanism than forward guidance. Forward guidance improves the situation at the end of the ZLB, which pulls up the economy during the entire ZLB. Government spending leaves the end of the ZLB unchanged: the economy reaches the natural steady state. Instead, government spending shifts the Phillips line upward, and with it, the field of trajectories. As a result, the natural steady state is connected to trajectories with higher consumption and inflation, which improves the situation during the entire ZLB (panel~A versus panel~B).

Just like forward guidance, government spending can have very strong effects in the NK model. When spending is low, the natural steady state is below the ZLB unstable line (panel~B). It is therefore connected to trajectories coming from the southwest quadrant of the phase diagram---just as without government spending (panel~A). Then, if the ZLB lasts longer, initial consumption and inflation fall lower. When spending is high enough that the unstable line crosses the natural steady state, the economy is also on the unstable line at time $0$ (panel~C). Finally, when spending is even higher, the natural steady state moves above the unstable line, so it is connected to trajectories coming from the northeast quadrant (panel~D). As a result, initial output and inflation are higher than previously. And as the ZLB lasts longer, initial output and inflation become even higher, without bound.

\begin{figure}[p]
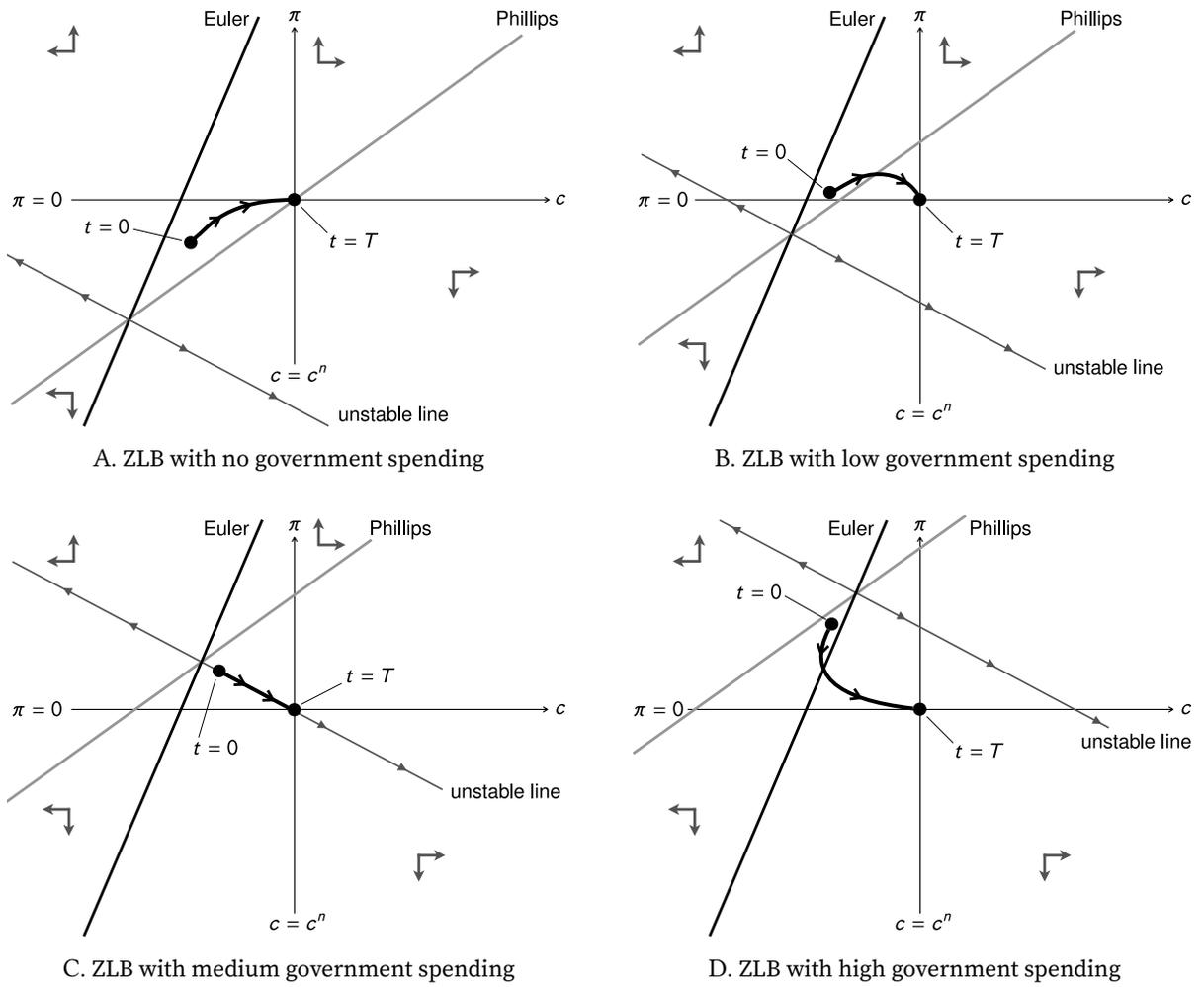

\subcaptionbox{ZLB with no government spending}{\includegraphics[scale=0.21,page=21]{\pdf}}\hfill
\subcaptionbox{ZLB with low government spending}{\includegraphics[scale=0.21,page=22]{\pdf}}\vspace{0.5cm}
\subcaptionbox{ZLB with medium government spending}{\includegraphics[scale=0.21,page=23]{\pdf}}\hfill
\subcaptionbox{ZLB with high government spending}{\includegraphics[scale=0.21,page=24]{\pdf}}
\caption{WUNK Model: ZLB Episodes with Government Spending}
\note{The figure describes various ZLB episodes with government spending in the WUNK model. The timeline of such episode is presented in table~\ref{t:scenarios}, panel~C. The panels display the phase diagrams of the linearized Euler-Phillips system for the WUNK model with government spending and convex disutility of labor at the ZLB: $c$ is private consumption; $\pi$ is inflation; $c^n$ is the natural level of private consumption; the Euler line is the locus $\dot{c}=0$; the Phillips line is the locus $\dot{\pi}=0$. The phase diagrams have the same properties as that in figure~\ref{f:phase}, panel~D, except that the Phillips line shifts upward when government spending increases (see equation \eqref{e:phillipsgss}). The equilibrium trajectory at the ZLB is the unique trajectory reaching the natural steady state at time~$T$. The four panels feature an increasing amount of government spending ($g$), starting from $g=0$ in panel~A. The figure shows that the NK model's anomaly disappears in the WUNK model: the government-spending multiplier is finite when the ZLB becomes arbitrarily long-lasting; and equilibrium trajectories are bounded irrespective of the duration of the ZLB.}
\label{f:spendingwunk}\end{figure}

The power of government spending at the ZLB is much weaker in the WUNK model (figure~\ref{f:spendingwunk}). Government spending does improves the situation at the ZLB, as inflation and consumption tend to be higher than without spending. But as the ZLB lasts longer, the position at the beginning of the ZLB converges to the ZLB steady state---unlike in the NK model, it does not go to infinity. So equilibrium trajectories are bounded, and government spending cannot generate unbounded booms.

Based on these dynamics, we isolate another anomaly in the NK model, which is resolved in the WUNK model (proof details in appendix~F):

\begin{proposition} Consider a ZLB episode during $(0,T)$, accompanied by government spending $g>0$. Let $c(t;g)$ and $y(t;g)$ be private consumption and output at time $t$; let $s>0$ be some incremental government spending; and let 
\begin{equation*}
m(g,s)=\frac{y(0;g+s/2)-y(0;g-s/2)}{s} = 1 + \frac{c(0;g+s/2)-c(0;g-s/2)}{s}
\end{equation*}
be the government-spending multiplier. 
\begin{itemize}
\item  In the NK model, there exists a government spending $g^*$ such that the government-spending multiplier becomes infinitely large when the ZLB duration approaches infinity: for any $s>0$, $\lim_{T\to \infty}m(g^*,s)=+\infty$. In addition, when government spending is above $g^*$, a long-enough ZLB episode generates an arbitrarily large boom: for any $g>g^*$, $\lim_{T\to \infty} c(0;g) = +\infty$.
\item In the WUNK model, in contrast, the multiplier has a finite limit when the ZLB duration approaches infinity: for any $g$ and $s$, when $T\to \infty$, $m(g,s)$ converges to
\begin{equation}
1 + \frac{\eta}{\frac{u'(0)\d\g a}{\e\k} \cdot \bp{\frac{\e}{\e-1}}^{\eta/(1+\eta)} - (1+\eta)}.
\label{e:multiplier}\end{equation}
Moreover, the economy is bounded above for any ZLB duration: let $c^g$ be private consumption in the ZLB steady state with government spending $g$; for any $T$ and for all $t\in(0,T)$, $c(t;g)<\max(c^g,c^n)$.
\end{itemize}\end{proposition}

The anomaly that a finite amount of government spending may generate an infinitely large boom as the ZLB becomes arbitrarily long-lasting is reminiscent of the findings by \ct[fig.~2]{CER09}, \ct[fig.~2]{W11}, and \ct[fig.~5]{C17}. They find that in the NK model government spending is exceedingly powerful when the ZLB is long-lasting.

In the WUNK model, this anomaly vanishes. \ct{DL17} and \ct{AD18} also obtain more realistic effects of government spending at the ZLB. In addition, \ct{BJS17} obtain moderate multipliers at the ZLB by introducing an endogenous liquidity premium in the New Keynesian model.

\section{Other New Keynesian Properties at the ZLB}

Beside the anomalous properties described in section~\ref{s:anomalies}, the New Keynesian model has several other intriguing properties at the ZLB: the paradoxes of thrift, toil, and flexibility; and a government-spending multiplier greater than one. We now show that the WUNK model shares these properties.

In the NK model these properties are studied in the context of a temporary ZLB episode. An advantage of the WUNK model is that we can simply work with a permanent ZLB episode. We assume that the natural rate of interest is permanently negative, and the central bank keeps the policy rate at zero forever. The only equilibrium is at the ZLB steady state, where the economy is in a slump: inflation is negative and output is below its natural level. The ZLB equilibrium is represented in figure~\ref{f:properties}: it is the intersection of a Phillips line, describing the steady-state Phillips curve, and an Euler line, describing the steady-state Euler equation. When an unexpected and permanent shock occurs, the economy jumps to a new ZLB steady state; we use the graphs to study such jumps.

\subsection{Paradox of Thrift}

We first study an increase in the marginal utility of wealth ($u'(0)$). The steady-state Phillips curve is unaffected, but the steady-state Euler equation changes. Using \eqref{e:rn}, we rewrite the steady-state Euler equation \eqref{e:eulerzlb}:
\begin{equation*}
\pi = -\d + \s +u'(0) y.
\end{equation*}
Increasing the marginal utility of wealth steepens the Euler line, which moves the economy inward along the Phillips line. Output and inflation therefore decrease (figure~\ref{f:properties}, panel~A). The following proposition gives the results:

\begin{proposition} At the ZLB in the WUNK model, the paradox of thrift holds: an unexpected and permanent increase in the marginal utility of wealth reduces output and inflation but does not affect relative wealth.\end{proposition}

The paradox of thrift was first discussed by Keynes, but it also appears in the New Keynesian model (\inp[p.~16]{E10}; \inp[p.~1486]{EK11}). When the marginal utility of wealth is higher, people want to increase their wealth holdings relative to their peers, so they favor saving over consumption. But in equilibrium, relative wealth is fixed at zero because everybody is the same; the only way to increase saving relative to consumption is to reduce consumption. In normal times, the central bank would offset this drop in aggregate demand by reducing nominal interest rates. This is not an option at the ZLB, so output falls.

\begin{figure}[p]
\subcaptionbox{Paradox of thrift}{\includegraphics[scale=0.21,page=25]{\pdf}}\hfill
\subcaptionbox{Paradox of toil}{\includegraphics[scale=0.21,page=26]{\pdf}}\vspace{0.5cm}
\subcaptionbox{Paradox of flexibility}{\includegraphics[scale=0.21,page=27]{\pdf}}\hfill
\subcaptionbox{Above-one government-spending multiplier}{\includegraphics[scale=0.21,page=28]{\pdf}}
\caption{WUNK model: Other Properties at the ZLB}
\note{The figure describes four comparative statics of the WUNK model at the ZLB. In panels~A, B, and C, the Euler and Phillips lines are the same as in figure~\ref{f:phase}, panel~D. In panel~D, the Euler and Phillips lines are the same as in figure~\ref{f:spendingwunk}. The ZLB equilibrium is at the intersection of the Euler and Phillips lines: output/consumption is below its natural level and inflation is negative. Panel~A illustrates the paradox of thrift: increasing the marginal utility of wealth steepens the Euler line, which depresses output and inflation without changing relative wealth. Panel~B illustrates the paradox of toil: reducing the disutility of labor moves the Phillips line outward, which depresses output, inflation, and hours worked. Panel~C illustrates the paradox of flexibility: decreasing the price-adjustment cost rotates the Phillips line counterclockwise around the natural steady state, which depresses output and inflation. Panel~D shows that the government-spending multiplier is above one: increasing government spending shifts the Phillips line upward, which raises private consumption and therefore increases output more than one-for-one.}
\label{f:properties}\end{figure}

\subsection{Paradox of Toil}

Next we consider a reduction in the disutility of labor ($\k$). In this case, the steady-state Phillips curve changes while the steady-state Euler equation does not. Using \eqref{e:yn}, we rewrite the steady-state Phillips curve \eqref{e:phillipsss}:
\begin{equation*}
\pi = \frac{\e\k}{\d\g a} y - \frac{\e-1}{\d\g}.
\end{equation*}
Reducing the disutility of labor flattens the Phillips line, which moves the economy inward along the Euler line. Thus, both output and inflation decrease (figure~\ref{f:properties}, panel~B). Since hours worked and output are related by $h=y/a$, hours fall as well. The following proposition states the results:

\begin{proposition} At the ZLB in the WUNK model, the paradox of toil holds: an unexpected and permanent reduction in the disutility of labor reduces output, inflation, and hours worked.\end{proposition}

The paradox of toil was discovered by \ct[p.~15]{E10} and \ct[p.~1487]{EK11}. It operates as follows. With lower disutility of labor, real marginal costs are lower, and the natural level of output is higher: producers would like to sell more. To increase sales, they reduce their prices by reducing inflation. At the ZLB, nominal interest rates are fixed, so the decrease in inflation raises real interest rates---which renders households more prone to save. In equilibrium, this lowers output and hours worked.\footnote{An increase in technology ($a$) would have the same effect as a reduction in the disutility of labor: it would lower output, inflation, and hours.}

\subsection{Paradox of Flexibility}

We then examine a decrease in the price-adjustment cost ($\g$). The steady-state Euler equation is not affected, but the steady-state Phillips curve is. Equation \eqref{e:phillipsss} shows that decreasing the price-adjustment cost leads to a counterclockwise rotation of the Phillips line around the natural steady state. This moves the economy downward along the Euler line, so output and inflation decrease (figure~\ref{f:properties}, panel~C). The following proposition records the results:

\begin{proposition} At the ZLB in the WUNK model, the paradox of flexibility holds: an unexpected and permanent decrease in price-adjustment cost reduces output and inflation.\end{proposition}

The paradox of flexibility was discovered by \ct[pp.~13--14]{W12} and \ct[pp.~1487--1488]{EK11}. Intuitively, with a lower price-adjustment cost, producers are keener to adjust their prices to bring production closer to the natural level of output. Since production is below the natural level at the ZLB, producers are keener to reduce their prices to stimulate sales. This accentuates the existing deflation, which translates into higher real interest rates. As a result, households are more prone to save, which in equilibrium depresses output.

\subsection{Above-One Government-Spending Multiplier}

We finally look at an increase in government spending ($g$), using the model with government spending introduced in section~4.3. From \eqref{e:phillipsgss} we see that increasing government spending shifts the Phillips line upward, which moves the economy upward along the Euler line: both private consumption and inflation increase (figure~\ref{f:properties}, panel~D). Since private consumption increases when public consumption does, the government-spending multiplier $dy/dg=1+dc/dg$ is greater than one. The ensuing proposition gives the results (proof details in appendix~F):

\begin{proposition} At the ZLB in the WUNK model, an unexpected and permanent increase in government spending raises private consumption and inflation. Hence the government-spending multiplier $dy/dg$ is above one; its value is given by \eqref{e:multiplier}.\end{proposition}

\ct{CER09}, \ct{E11}, and \ct{W11} also show that at the ZLB in the New Keynesian model, the government-spending multiplier is above one. The intuition is the following. With higher government spending, real marginal costs are higher for a given level of sales to households. Producers pass the cost increase through into prices, which raises inflation. At the ZLB, the increase in inflation lowers real interest rates---as nominal interest rates are fixed---which deters households from saving. In equilibrium, this leads to higher private consumption and a multiplier above one.

\section{Empirical Assessment of the WUNK Assumption}

In the WUNK model, the marginal utility of wealth is assumed to be high enough that the steady-state Euler equation is steeper than the steady-state Phillips curve at the ZLB. We assess this assumption using US evidence.

As a first step, we re-express the WUNK assumption in terms of estimable statistics. We obtain the following condition (derivations in appendix~G):
\begin{equation}
\d - r^n  > \frac{\l}{\d},
\label{e:wunk2}\end{equation} 
where $\d$ is the time discount rate, $r^n$ is the average natural rate of interest, and $\l$ is the coefficient on output gap in a New Keynesian Phillips curve. The term $\d - r^n$ measures the marginal rate of substitution between wealth and consumption, $u'(0) y^n$. It indicates how high the marginal utility of wealth is and thus how steep the steady-state Euler equation is at the ZLB. The term $\l/\d$ indicates how steep the steady-state Phillips curve is. The $\d$ comes from the denominator of the slopes of the Phillips curves \eqref{e:phillipsss} and \eqref{e:phillipsgss}; the $\l$ measures the rest of the slope coefficients. Condition \eqref{e:wunk2} is expressed in terms of sufficient statistics, so it applies both when the disutility of labor is linear (in which case it is equivalent to \eqref{e:wunk}) and when the disutility of labor is convex (in which case it is equivalent to \eqref{e:wunkg}). We now survey the literature to obtain estimates of $r^n$, $\l$, and $\d$.

\subsection{Natural Rate of Interest}

A large number of macroeconometric studies have estimated the natural rate of interest, using different statistical models, methodologies, and data. Recent studies obtain comparable estimates of the natural rate for the United States: around $2\%$ per annum on average between 1985 and 2015 \cp[fig.~1]{W17}. Accordingly, we use $r^n=2\%$ as our estimate.

\subsection{Output-Gap Coefficient in the New Keynesian Phillips Curve}

Many studies have estimated New Keynesian Phillips curves. \ct[sec.~5]{MPS14} offer a synthesis for the United States. They generate estimates of the New Keynesian Phillips curve using an array of US data, methods, and specifications found in the literature. They find significant uncertainty around the estimates, but in many cases the output-gap coefficient is positive and very small. Overall, their median estimate of the output-gap coefficient is $\l=0.004$ (table~5, row~1), which we use as our estimate.

\subsection{Time Discount Rate}

Since the 1970s, many studies have estimated time discount rates using field and laboratory experiments and real-world behavior. \ct[table~1]{FLO02} survey 43 such studies. The estimates are quite dispersed, but the majority of them points to high discount rates, much higher than prevailing market interest rates. We compute the mean estimate in each of the studies covered by the survey, and then compute the median value of these means. We obtain an annual discount rate of $\d=35\%$. 

There is one immediate limitation with the studies discussed by \name{FLO02}: they use a single rate to exponentially discount future utility. But exponential discounting does not describe reality well because people seem to choose more impatiently for the present than for the future---they exhibit present-focused preferences \cp{EL19}. Recent studies have moved away from exponential discounting and allowed for present-focused preferences, including quasi-hyperbolic ($\b$-$\d$) discounting. \ct[table~3]{AHL14} survey 16 such studies, concentrating on experimental studies with real incentives. We compute the mean estimate in each study and then the median value of these means; we obtain an annual discount rate of $\d=43\%$. Accordingly, even after accounting for present-focus, time discounting remains high. We use $\d=43\%$ as our estimate.\footnote{There are two potential issues with the experiments discussed in \ct{AHL14}. First, many are run with university students instead of subjects representative of the general population. There does not seem to be systematic differences in discounting between student and non-student subjects, however \cp[sec.~6A]{CEL19}. Hence, using students is unlikely to bias the estimates reported by \name{AHL14}. Second, all the experiments elicit discount rates using financial flows, not consumption flows. As the goal is to elicit the discount rate on consumption, this could be problematic \cp[sec.~4B]{CEL19}; the problems could be exacerbated if subjects derive utility from wealth. To assess this potential issue, suppose first (as in most of the literature) that monetary payments are consumed at the time of receipt, and that the utility function is locally linear. Then the experiments deliver estimates of the relevant discount rate \cp[sec.~4B]{CEL19}. If these conditions do not hold, the experimental findings are more difficult to interpret. For instance, if subjects optimally smooth their consumption over time by borrowing and saving, then the experiments only elicit the interest rate faced by subjects, and reveal nothing about their discount rate \cp[sec.~4B]{CEL19}. In that case, we should rely on experiments using time-dated consumption rewards instead of monetary rewards. Such experiments directly deliver estimates of the discount rate. Many such experiments have been conducted; a robust finding is that discount rates are systematically higher for consumption rewards than for monetary rewards \cp[sec.~3A]{CEL19}. Hence, the estimates presented in \name{AHL14} are, if anything, lower bounds on actual discount rates.}

\subsection{Assessment}

We now combine our estimates of $r^n$, $\l$, and $\d$ to assess the WUNK assumption. Since $\l$ is estimated using quarters as units of time, we re-express $r^n$ and $\d$ as quarterly rates: $r^n = 2\% / 4 = 0.5\%$ per quarter, and $\d = 43\% / 4 = 10.8\% $ per quarter. We conclude that \eqref{e:wunk2} comfortably holds: $\d-r^n = 0.108-0.005= 0.103$, which is much larger than $\l/\d =0.004/0.108=0.037$. Hence the WUNK assumption holds in US data. 

The discount rate used here (43\% per annum) is much higher than discount rates used in macroeconomic models (typically less than 5\% per annum). This is because our discount rate is calibrated from microevidence, while the discount rate in macroeconomic models is calibrated to match observed real interest rates. 

This discrepancy occasions two remarks. First, the wealth-in-the-utility assumption is advantageous because it accords with the fact that people exhibit double-digit time discount rates and yet are willing to save at single-digit interest rates. In the standard model, by contrast, the discount rate necessarily equals the real interest rate in steady state, so the model cannot have $\d\gg 5\%$.

Second, the WUNK assumption would also hold with discount rates below 43\%. Indeed, \eqref{e:wunk2} holds for discount rates as low as 27\% because $\d-r^n = (0.27/4)-0.005= 0.062$ is greater than $\l/\d = 0.004/(0.27/4) = 0.059$. An annual discount rate of 27\% is at the low end of available microestimates: in 11 of the 16 studies in \ct[table~3]{AHL14}, the bottom of the estimate range is above 27\%; and in 13 of the 16 studies, the mean estimate is above 27\%.

Finally, while our model omits firms and assumes that households are both producers and consumers, in reality firms and households are often separate entities that could have different discount rates. With different discount rates, \eqref{e:wunk2} would become
\begin{equation*}
\d^h-r^n > \frac{\l}{\d^f},
\end{equation*}
where $\d^h$ is households' discount rate and $\d^f$ is firms' discount rate. Clearly, if firms have a low discount rate, the WUNK assumption is less likely to be satisfied. If we use $\d^h=43\%$, $r^n=2\%$, and $\l=0.004$, we find that the WUNK condition holds as long as firms have an annual discount rate above 16\% because $\d^h-r^n = (0.43/4)-0.005= 0.103$ is greater than $\l/\d^f = 0.004/(0.16/4) = 0.100$. A discount rate of 16\% is only slightly above that reported by large US firms: in a survey of 228 CEOs, \ct{PS95} find an average annual real discount rate of 12.2\%; and in a survey of 86 CFOs, \ct[p.~447]{JMM16} find an average annual real discount rate of 12.7\%.

\section{Conclusion}

This paper proposes an extension of the New Keynesian model that is immune to the anomalies that plague the standard model at the ZLB. The extended model deviates only minimally from the standard model: relative wealth enters the utility function, which only adds an extra term in the Euler equation. Yet, when the marginal utility of wealth is sufficiently high, the model behaves well at the ZLB: even when the ZLB is long-lasting, there is no collapse of inflation and output, and both forward guidance and government spending have limited, plausible effects. The extended model also retains other properties of the standard model at the ZLB: the paradoxes of thrift, toil, and flexibility; and a government-spending multiplier greater than one.

Our analysis would apply more generally to any New Keynesian model representable by a discounted Euler equation and a Phillips curve \eg{DGP12,G16,MNS17,CFJ17,BP18,AL16}. Wealth in the utility function is a simple way to generate discounting; but any model with discounting would have similar phase diagrams and properties. Hence, for such models to behave well at the ZLB, there is only one requirement: that discounting is strong enough to make the steady-state Euler equation steeper than the steady-state Phillips curve at the ZLB; the source of discounting is unimportant. In the real world, several discounting mechanisms might operate at the same time and reinforce each other. A model blending these mechanisms would be even more likely to behave well at the ZLB.

\bibliography{\bib}

\appendix

\section{Formal derivation of Euler equation \& Phillips curve}\label{a:formal}

We derive the two differential equations that describe the equilibrium of the New Keynesian model with wealth in the utility function: the Phillips curve, given by (1); and the Euler equation, given by (4).

\subsection{Household's problem}\label{a:hamiltonian}

We begin by solving household $j$'s problem. The current-value Hamiltonian of the problem is
\begin{align*}
\Hc_j &= \frac{\e}{\e-1} \ln{\int_{0}^{1} c_{jk}(t)^{(\e-1)/\e}\,dk} + u\of{\frac{b_{j}(t)-b(t)}{p(t)}} -\frac{\k}{a}y^d_{j}(p_{j}(t),t)-\frac{\g}{2} \pi_{j}(t)^2 \\
& + \Ac_{j}(t) \bs{i^h(t) b_{j}(t) + p_{j}(t)  y^d_{j}(p_{j}(t),t) - \int_0^1 p_{k}(t) c_{jk}(t)\,dk - \tau(t)}+\Bc_{j}(t) \pi_{j}(t) p_{j}(t),
\end{align*}
with control variables $c_{jk}(t)$ for all $k\in[0,1]$ and $\pi_{j}(t)$, state variables $b_{j}(t)$ and $p_{j}(t)$, and costate variables $\Ac_{j}(t)$ and $\Bc_{j}(t)$. Note that we have used the production and demand constraints to substitute $y_{j}(t)$ and $h_{j}(t)$ out of the Hamiltonian. (To ease notation we now drop the time index~$t$.)

We apply the necessary conditions for a maximum to the household's problem given by \ct[theorem~7.9]{A09}. These conditions form the basis of the model's equilibrium conditions.

The first optimality conditions are $\pdx{\Hc_j}{c_{jk}}=0$ for all $k\in [0,1]$. They yield
\begin{equation}
\frac{1}{c_{j}} \bp{\frac{c_{jk}}{c_{j}}}^{-1/\e}= \Ac_{j}  p_{k}.
\label{e:aj0}\end{equation}
Appropriately integrating \eqref{e:aj0} over all $k\in [0,1]$ and using the expressions for the consumption and price indices,
\begin{align}
c_{j}(t) &= \bs{\int_{0}^{1} c_{jk}(t)^{(\e-1)/\e}\,dk}^{\e/(\e-1)}\label{e:cj}\\
p(t) &= \bs{\int_{0}^1 p_{j}(t)^{1-\e}\,di}^{1/(1-\e)},\label{e:p}
\end{align}
we find
\begin{equation}
\Ac_{j} =\frac{1}{p c_{j}}.
\label{e:aj}\end{equation}
Moreover, combining \eqref{e:aj0} and \eqref{e:aj}, we obtain
\begin{equation}
c_{jk}= \bp{\frac{p_k}{p}}^{-\e} c_j. 
\label{e:cjk}\end{equation}
Integrating \eqref{e:cjk} over all $j\in [0,1]$, we get the usual demand for good $k$:
\begin{equation}
y^d_{k}(p_{k}) = \int_0^1 c_{jk}\,dj = \bp{\frac{p_{k}}{p}}^{-\e} c,
\label{e:ydk}\end{equation}
where $c = \int_0^1 c_{j}\,dj$ is aggregate consumption. We use this expression for $y^d_{k}(p_{k})$ in household~$k$'s Hamiltonian. Equation \eqref{e:cjk} also implies that
\begin{equation*}
\int_0^1 p_{k} c_{jk}\,dk = \int_0^1 p_{k} \bp{\frac{p_k}{p}}^{-\e} c_j \,dk = p c_{j}.
\end{equation*}
This means that when consumption expenditure is allocated optimally across goods, the price of one unit of consumption index is $p$.

The second optimality condition is $\pdx{\Hc_j}{b_{j}}=\d\Ac_{j}-\dot{\Ac}_{j}$, which gives
\begin{equation*}
-\frac{\dot{\Ac}_{j}}{\Ac_{j}}=i^h+\frac{1}{p \Ac_{j}}\cdot u'\of{\frac{b_{j}-b}{p}}-\d.
\end{equation*}
Using \eqref{e:aj} and $i^h = i + \s$, we obtain the household's Euler equation:
\begin{equation}
\frac{\dot{c}_{j}}{c_{j}}= i + \s - \pi +  c_{j} u'\of{\frac{b_{j}-b}{p}} -\d.
\label{e:eulerj}\end{equation}
This equation describes the optimal path for household $j$'s consumption.

The third optimality condition is $\pdx{\Hc_j}{\pi_{j}}=0$, which yields
\begin{equation}
\Bc_{j} p_{j} = \g\pi_{j}.
\label{e:bj}\end{equation}
Differentiating \eqref{e:bj} with respect to time, we obtain
\begin{equation}
\frac{\dot{\Bc}_{j}}{\Bc_{j}} = \frac{\dot{\pi}_{j}}{\pi_{j}} - \pi_{j}.
\label{e:bjdot}\end{equation}

The last optimality condition is $\pdx{\Hc_j}{p_{j}}=\d \Bc_{j}-\dot{\Bc}_{j}$, which implies
\begin{equation*}
\frac{\k}{a} \cdot \frac{\e y_{j}}{p_{j}} - (\e-1) \Ac_{j} y_{j} + \Bc_{j} \pi_{j} = \d \Bc_{j}-\dot{\Bc}_{j}.
\end{equation*}
Reshuffling the terms then yields
\begin{equation*}
\pi_{j} - \frac{(\e-1) y_{j} \Ac_{j}}{\Bc_{j} p_{j}} \bp{p_{j} -\frac{\e}{\e-1}\cdot \frac{\k}{a\Ac_{j}}} = \d-\frac{\dot{\Bc}_{j}}{\Bc_{j}}.
\end{equation*}
Finally, incorporating \eqref{e:aj}, \eqref{e:bj}, and \eqref{e:bjdot}, we obtain the household's Phillips curve:
\begin{equation}
\frac{\dot{\pi}_{j}}{\pi_{j}} = \d + \frac{(\e-1) y_{j}}{\g c_{j} \pi_{j}} \bp{\frac{p_{j}}{p}-\frac{\e}{\e-1}\cdot\frac{\k c_{j}}{a}}.
\label{e:phillipsj}\end{equation}
This equation describes the optimal path for the price set by household $j$.

\subsection{Equilibrium}

We now describe the equilibrium of the model. Since all households face the same initial conditions, they all behave the same. We therefore drop the subscripts~$j$ and $k$ on all the variables. In particular, all households hold the same wealth, so relative wealth is zero: $b_j=b$. In addition, production and consumption are equal in equilibrium: $y=c$.

Accordingly, the household's Phillips curve, given by \eqref{e:phillipsj}, simplifies to
\begin{equation*}
\dot{\pi} = \d \pi -  \frac{\e\k}{\g a}  \bp{y-y^n},
\end{equation*}
where
\begin{equation}
y^n= \frac{\e-1}{\e}\cdot \frac{a}{\k}. 
\label{e:yna}\end{equation}
And the household's Euler equation, given by \eqref{e:eulerj}, simplifies to
\begin{equation*}
\frac{\dot{y}}{y}= r - r^n + u'(0) (y-y^n),
\end{equation*} 
where $r=i-\pi$ and
\begin{equation}
r^n = \d - \s - u'(0)y^n. 
\label{e:rna}\end{equation}
These differential equations are the Phillips curve (1) and Euler equation (4).

\section{Heuristic derivation of Euler equation \& Phillips curve}
		
To better understand and interpret the continuous-time Euler equation and Phillips curve, we complement the formal derivations of appendix~\ref{a:formal} with heuristic derivations, as in \ct[pp.~40--42]{BF89}.

\subsection{Euler equation}

The Euler equation says that households save in an optimal fashion: they cannot improve their situation by shifting consumption a little bit across time. 

Consider a household delaying consumption of one unit of output from time $t$ to time $t+dt$. The unit of output, invested at a real interest rate $r^h(t)$, becomes $1 +r^h(t) dt$ at time $t+dt$. Given log consumption utility, the marginal utility from consumption at any time $t$ is $e^{-\d t}/y(t)$. Hence, the household forgoes $e^{-\d t}/y(t)$ utils at time $t$ and gains 
\begin{equation*}
[1 +r^h(t) dt] \frac{e^{-\d (t+dt)}}{y(t+dt)}
\end{equation*}
utils at time $t+dt$.

Since people enjoy holding wealth, the one unit of output saved between $t$ and $t+dt$ provides hedonic returns in addition to financial returns. The marginal utility from real wealth at time $t$ is $e^{-\d t} u'(0)$. Hence, by holding an extra unit of real wealth for a duration $dt$, the household gains $e^{-\d t} u'(0) dt$ utils. 

At the optimum, reallocating consumption over time does not affect utility, so the following holds:
\begin{equation*}
0 =  -   \frac{e^{-\d t}}{y(t)} +  \bs{1 +r^h(t) dt} \frac{e^{-\d (t+dt)}}{y(t+dt)} + e^{-\d t} u'(0) dt.
\end{equation*}
Divided by $e^{-\d t}/y(t)$, this condition becomes
\begin{equation*}
1 =[1 +r^h(t) dt]e^{-\d dt}  \frac{y(t)}{y(t+dt)} + u'(0) y(t) dt.
\end{equation*}
Furthermore, up to second-order terms, the following approximations are valid: 
\begin{align*}
e^{-\d dt}&= 1 - \d dt\\
\frac{y(t+dt)}{y(t)}&= 1 + \frac{\dot{y}(t)}{y(t)} dt\\
\frac{1}{1+x dt}&= 1- x dt,\quad\text{for any }x.
\end{align*}
Hence, up to second-order terms, the previous condition gives
\begin{equation*}
1 =  \bs{1 +r^h(t) dt} \bp{1 - \d dt} \bs{1 - \frac{\dot{y}(t)}{y(t)} dt}+ u'(0) y(t) dt.
\end{equation*}
Keeping only first-order terms, we obtain
\begin{equation*}
 1 = 1 - \d dt + r^h(t) dt - \frac{\dot{y}(t)}{y(t)} dt + u'(0) y(t) dt.
\end{equation*}
Reshuffling the terms and dividing by $dt$, we conclude that
\begin{equation*}
\frac{\dot{y}(t)}{y(t)} = r^h(t) -\d + u'(0) y(t).
\end{equation*}
We obtain the Euler equation (4) from here by noting that $r^h(t) = r(t) + \s$ and introducing the natural rate of interest $r^n$ given by \eqref{e:rna}.

\subsection{Phillips curve}

The Phillips curve says that households price in an optimal fashion: they cannot improve their situation by shifting inflation a little bit across time.

Consider a household delaying one percentage point of inflation from time $t$ to time $t+dt$. Given the quadratic price-change disutility, the marginal disutility from inflation at any time $t$ is $e^{-\d t} \g \pi(t)$. Hence, at time $t$, the household avoids a disutility of 
\begin{equation*}
e^{-\d t} \g \pi(t) \times 1\%.
\end{equation*}
And, at time $t+dt$, the household incurs an extra disutility of 
\begin{equation*}
e^{-\d (t+dt)} \g \pi(t+dt) \times 1\%.
\end{equation*}

Delaying inflation by one percentage point reduces the household's price between times $t$ and $t+dt$ by $dp(t) = -1\% \times p(t)$. The price drop then affects sales. Since the price elasticity of demand is $-\e$, sales increase by 
\begin{equation*}
dy(t) = -\e y(t) \times -1\%  = \e y(t)\times 1\%.
\end{equation*}
Accordingly, the household's revenue grows by 
\begin{equation*}
d(p(t) y(t)) = p(t) dy(t) + y(t)dp(t)  = (\e-1)y(t)p(t)\times 1\%.
\end{equation*}

With a higher revenue, the household can afford to consume more. Since in equilibrium all prices are the same, equal to $p(t)$, the increase in revenue raises consumption by 
\begin{equation*}
dc(t)=\frac{d(p(t) y(t))}{p(t)} = (\e-1) y(t) \times 1\%. 
\end{equation*}
Hence, between times $t$ and $t+dt$, the utility of consumption increases by 
\begin{equation*}
\frac{e^{-\d t}}{y(t)} dc(t) =e^{-\d t}(\e-1) \times 1\%.
\end{equation*}

At the same time, because production is higher, the household must work more. Hours worked are extended by 
\begin{equation*}
dh(t) = \frac{dy(t)}{a} = \frac{\e y(t)}{a} \times 1\%. 
\end{equation*}
As a result, between times $t$ and $t+dt$, the disutility of labor is elevated by 
\begin{equation*}
e^{-\d t} \k dh(t) = e^{-\d t} \frac{\k\e y(t)}{a} \times 1\%.
\end{equation*}

At the optimum, shifting inflation across time does not affect utility, so the following holds:
\begin{equation*}
0 = e^{-\d t} \g  \pi(t) \times 1\% - e^{-\d (t+dt)} \g \pi(t+dt) \times 1\% + e^{-\d t}  (\e-1) \times 1\%\times dt- e^{-\d t}\k\e \frac{y(t)}{a} \times 1\% \times dt.
\end{equation*}
Divided by $e^{-\d t}\times 1\%$, this condition yields
\begin{equation*}
0 = \g  \pi(t) - e^{-\d dt} \g \pi(t+dt) + (\e-1) \times dt- \k\e \frac{y(t)}{a} \times dt.
\end{equation*}
Furthermore, up to second-order terms, the following approximations hold:
\begin{align*}
e^{-\d dt} &= 1 - \d dt\\
\pi(t+dt) &= \pi(t) + \dot{\pi}(t) dt.
\end{align*}
Therefore, up to second-order terms, the previous condition gives
\begin{equation*}
0 = \g  \pi(t) - (1 - \d dt) \g \bs{\pi(t) + \dot{\pi}(t) dt} - \k\e \frac{y(t)}{a} dt + (\e-1) dt
\end{equation*}
Then, keeping only first-order terms, we obtain
\begin{equation*}
0 = \d\g \pi(t) dt -\g\dot{\pi}(t) dt - \k\e \frac{y(t)}{a} dt + (\e-1)dt.
\end{equation*}
Rearranging the terms and dividing by $\g dt$, we conclude that
\begin{equation*}
\dot{\pi}(t) = \d  \pi(t) - \frac{\e\k}{\g a}\bs{y(t) - \frac{\e-1}{\e}\cdot \frac{a}{\k}}.
\end{equation*}
Once we introduce the natural level of output $y^n$ given by \eqref{e:yna}, we obtain the Phillips curve (1).

The Phillips curve implies that without price-adjustment cost ($\g=0$), households would produce at the natural level of output. This result comes from the monopolistic nature of competition. Without price-adjustment cost, it is optimal to charge a relative price that is a markup $\e/(\e-1)$ over the real marginal cost. In turn, the real marginal cost is the marginal rate of substitution between labor and consumption divided by the marginal product of labor. In equilibrium, all relative prices are $1$, the marginal rate of substitution between labor and consumption is $\k/(1/y)=\k y$, and the marginal product of labor is $a$. Hence, optimal pricing requires
\begin{equation*}
1 = \frac{\e}{\e-1}\cdot \frac{\k y}{a}.
\end{equation*}
Combined with \eqref{e:yna}, this condition implies $y=y^n$.

The derivation also elucidates why in steady state, inflation is positive whenever output is above its natural level. When inflation is positive, a household can reduce its price-adjustment cost by lowering its inflation. Since pricing is optimal, however, there cannot exist any profitable deviation from the equilibrium. This means that the household must also incur a cost when it lowers inflation. A consequence of lowering inflation is that the price charged by the household drops, which stimulates its sales and production. The absence of profitable deviation imposes that the household incurs a cost when production increases. In other words, production must be excessive: output must be above its natural level.

\section{Euler equation \& Phillips curve in discrete time}

We recast the model of section~3 in discrete time, and we rederive the Euler equation and Phillips curve. This reformulation might be helpful to compare our model to the textbook New Keynesian model, which is presented in discrete time \cp{Woo03,G08}. The reformulation also shows that introducing wealth in the utility function yields a discounted Euler equation.

\subsection{Assumptions}

The assumptions are the same in the discrete-time model as in the continuous-time model, except for government bonds. In discrete time, households trade one-period government bonds. Bonds purchased in period~$t$ have a price $q(t)$ and pay one unit of money in period~$t+1$. The nominal interest rate on government bonds is defined as $i^h(t) = -\ln(q(t))$.

\subsection{Household's problem}

Household $j$ chooses sequences $\bc{y_{j}(t), p_{j}(t), h_{j}(t), \bs{c_{jk}(t)}_{k=0}^{1}, b_{j}(t)}_{t=0}^{\infty}$ to maximize the discounted sum of instantaneous utilities
\begin{equation*}
\sum_{t=0}^{\infty}\b^{t}  \bc{\frac{\e}{\e-1} \ln{\int_{0}^{1} c_{jk}(t)^{(\e-1)/\e}\,dk} +u\of{\frac{b_{j}(t)-b(t)}{p(t)}}-\k h_{j}(t)-\frac{\g}{2} \bs{\frac{p_j(t)}{p_j(t-1)}-1}^2}dt,
\end{equation*}
where $\b<1$ is the time discount factor. The maximization is subject to three constraints. First, there is a production function: $y_{j}(t) = a h_{j}(t)$. Second, there is the demand for good~$j$:
\begin{equation*}
y_{j}(t) = \bs{\frac{p_{j}(t)}{p(t)}}^{-\e} c(t) \equiv y^d_{j}(p_{j}(t),t).
\end{equation*}
The demand for good $j$ is the same as in continuous time because the allocation of consumption expenditure across goods is a static decision, so it is unaffected by the representation of time. And third, there is a budget constraint:
\begin{equation*}
\int_{0}^{1} p_{k}(t)  c_{jk}(t)\,dk + q(t)  b_{j}(t) + \tau(t) = p_{j}(t) y_{j}(t)+ b_{j}(t-1).
\end{equation*}
Household~$j$ is also subject to a solvency constraint preventing Ponzi schemes. Lastly, household~$j$ takes as given the initial conditions $b_{j}(-1)$ and $p_j(-1)$, as well as the sequences of aggregate variables $\bc{p(t), q(t), c(t)}_{t=0}^{\infty}$.

The Lagrangian of the household's problem is
\begin{align*}
\Lc_j & =\sum_{t=0}^{\infty}\b^{t}  \bigg\{\frac{\e}{\e-1} \ln{\int_{0}^{1} c_{jk}(t)^{(\e-1)/\e}\,dk} +u\of{\frac{b_{j}(t)-b(t)}{p(t)}}- \frac{\k}{a}y^d_{j}(p_{j}(t),t)-\frac{\g}{2} \bs{\frac{p_j(t)}{p_j(t-1)}-1}^2 \\
&+\Ac_{j}(t)  \bs{p_{j}(t) y^d_{j}(p_{j}(t),t)+ b_{j}(t-1) - \int_{0}^{1} p_{k}(t)  c_{jk}(t)\,dk - q(t)  b_{j}(t) - \tau(t)}\bigg\}
\end{align*}
where $\Ac_{j}(t)$ is a Lagrange multiplier. We have used the production and demand constraints to substitute $h_{j}(t)$ and $y_{j}(t)$ out of the Lagrangian. 

The necessary conditions for a maximum to the household's problem are standard first-order conditions. The first optimality conditions are $\pdx{\Lc_j}{c_{jk}(t)}=0$ for all $k\in [0,1]$ and all $t$. As in continuous time, these conditions yield
\begin{equation}
\Ac_{j}(t) =\frac{1}{p(t) c_{j}(t)}.
\label{e:ajd}\end{equation}

The second optimality condition is $\pdx{\Lc_j}{b_{j}(t)}=0$ for all $t$, which gives
\begin{equation*}
q(t)  \Ac_{j}(t) = \frac{1}{p(t)} u'\of{\frac{b_{j}(t)-b(t)}{p(t)}} + \b \Ac_{j}(t+1).
\end{equation*} 
Using \eqref{e:ajd}, we obtain the household's Euler equation:
\begin{equation}
q(t)= c_j(t) u'\of{\frac{b_{j}(t)-b(t)}{p(t)}}+ \b  \frac{p(t)  c_{j}(t)}{p(t+1)  c_{j}(t+1)}.
\label{e:eulerjd}\end{equation}

The third optimality condition is $\pdx{\Lc_j}{p_{j}(t)}=0$ for all $t$, which yields
\begin{equation*}
0=\frac{\k}{a}\cdot\frac{\e y_j(t)}{p_j(t)}-\frac{\g}{p_j(t-1)}\bs{\frac{p_j(t)}{p_j(t-1)}-1} + (1-\e) \Ac_{j}(t) y_{j}(t)+\b\g \frac{p_j(t+1)}{p_j(t)^2}\bs{\frac{p_j(t+1)}{p_j(t)}-1}.
\end{equation*}
Multiplying this equation by $p_j(t)/\g$ and using \eqref{e:ajd}, we obtain the household's Phillips curve:
\begin{equation}
\frac{p_j(t)}{p_j(t-1)}\bs{\frac{p_j(t)}{p_j(t-1)}-1} = \b \frac{p_j(t+1)}{p_j(t)}\bs{\frac{p_j(t+1)}{p_j(t)}-1} + \frac{\e\k}{\g a} y_j(t)-\frac{\e-1}{\g} \cdot \frac{p_{j}(t) y_{j}(t)}{p(t)c_j(t)}.
\label{e:phillipsjd}\end{equation}

\subsection{Equilibrium}

We now describe the equilibrium. Since all households face the same initial conditions, they all behave the same, so we drop the subscripts~$j$ and $k$ on all the variables. In particular, all households hold the same wealth, so relative wealth is zero: $b_j(t)=b(t)$. In addition, production and consumption are equal in equilibrium: $y(t)=c(t)$. 

Accordingly, from \eqref{e:eulerjd} we obtain the Euler equation
\begin{equation}
q(t)=u'(0) y(t)+ \b \frac{p(t)  y(t)}{p(t+1)  y(t+1)}.
\label{e:eulerdnl}\end{equation}
Moreover, combining \eqref{e:phillipsjd} and \eqref{e:yna}, we obtain the Phillips curve
\begin{equation}
\frac{p(t)}{p(t-1)}\bs{\frac{p(t)}{p(t-1)}-1}=\b \frac{p(t+1)}{p(t)}\bs{\frac{p(t+1)}{p(t)}-1} + \frac{\e-1}{\g} \bs{\frac{y(t)}{y^n}-1}.
\label{e:phillipsdnl}\end{equation}

\subsection{Log-linearization}
		
To obtain the standard expressions of the Euler equation and Phillips curve, we log-linearize \eqref{e:eulerdnl} and \eqref{e:phillipsdnl} around the natural steady state: where $y=y^n$, $\pi=0$, and $i=r^n$. To that end, we introduce the log-deviation of output from its steady-state level: $\hat{y}(t) = \ln(y(t)) - \ln(y^n)$. We also introduce the inflation rate between periods~$t$ and $t+1$: $\pi(t+1) = \ln(p(t+1))-\ln(p(t))$.

\paragraph{Euler equation.} We start by log-linearizing the Euler equation \eqref{e:eulerdnl}. 

We first take the log of the left-hand side of \eqref{e:eulerdnl}. Using the discrete-time definition of the nominal interest rate faced by households, $i^h(t)$, we obtain $\ln(q(t))=-i^h(t)$. At the natural steady state, the monetary-policy rate is $i = r^n$, so the interest rate faced by households is $i^h = r^n + \s$, and $\ln(q(t))=-r^n - \s$.

Next we take the log of the right-hand side of \eqref{e:eulerdnl}. We obtain $\L\equiv\ln(\L_1+\L_2)$, where 
\begin{equation*}
\L_1 \equiv u'(0) y(t),\qquad \L_2 \equiv \b \frac{p(t)  y(t)}{p(t+1)  y(t+1)}.
\end{equation*}
For future reference, we compute the values of $\L$, $\L_1$, and $\L_2$ at the natural steady state. At the natural steady state, the log of the left-hand side of \eqref{e:eulerdnl} equals $-r^n-\s$, which implies that the log of the right-hand side of \eqref{e:eulerdnl} must also equal $-r^n-\s$. That is, at the natural steady state, $\L=-r^n-\s$. Moreover, at that steady state, $\L_1=u'(0)y^n$. And, since inflation is zero and output is constant at that steady state, $\L_2 = \b$.

Using these results, we obtain a first-order approximation of $\L(\L_1,\L_2)$ around the natural steady state:
\begin{equation*}
\L = -r^n-\s + \pd{\L}{\L_1}\bs{\L_1-u'(0)y^n} + \pd{\L}{\L_2}\bs{\L_2-\b}.
\end{equation*}
Factoring out $u'(0)y^n$ and $\b$, and using the definitions of $\L_1$ and $\L_2$, we obtain
\begin{equation}
\L = -r^n-\s + u'(0)y^n \cdot \pd{\L}{\L_1} \cdot \bs{\frac{y(t)}{y^n}-1} + \b\cdot\pd{\L}{\L_2}\cdot \bs{\frac{p(t)  y(t)}{p(t+1)  y(t+1)}-1}.
\label{e:foa1}\end{equation}
Since $\L=\ln(\L_1+\L_2)$, we obviously have
\begin{equation*}
\pd{\L}{\L_1}=\pd{\L}{\L_2}=\frac{1}{\L_1+\L_2}.
\end{equation*}
In \eqref{e:foa1}, the derivatives are evaluated at the natural state, so
\begin{equation*}
\pd{\L}{\L_1}=\pd{\L}{\L_2}=\frac{1}{u'(0)y^n + \b}.
\end{equation*}
Hence, \eqref{e:foa1} becomes
\begin{equation}
\L = -r^n-\s + \frac{u'(0)y^n}{u'(0)y^n + \b} \bs{\frac{y(t)}{y^n}-1} + \frac{\b}{u'(0)y^n + \b} \bs{\frac{p(t)  y(t)}{p(t+1)  y(t+1)}-1}.
\label{e:foa2}\end{equation}

Last, up to second-order terms, we have $\ln(x)=x-1$ around $x=1$. Thus, we have the following first-order approximations around the natural steady state:
\begin{equation}
\frac{y(t)}{y^n}-1 =\ln{\frac{y(t)}{y^n}} = \hat{y}(t) 
\label{e:ytyn}\end{equation}
and
\begin{align*}
\frac{p(t)  y(t)}{p(t+1)  y(t+1)}-1 &= \ln{\frac{p(t)  y(t)}{p(t+1)  y(t+1)}}\\
&= \ln{\frac{y(t)}{y^n}} - \ln{\frac{y(t+1)}{y^n}}-\ln{\frac{p(t+1)}{p(t)}} \\
&= \hat{y}(t)-\hat{y}(t+1)-\pi(t+1).
\end{align*}
We can therefore rewrite \eqref{e:foa2} as
\begin{equation*}
\L= -r^n-\s + \frac{u'(0)y^n}{u'(0)y^n + \b} \hat{y}(t) +  \frac{\b}{u'(0)y^n + \b} \bs{\hat{y}(t)-\hat{y}(t+1)-\pi(t+1)}.
\end{equation*}
Finally, introducing 
\begin{equation*}
\a = \frac{\b}{\b + u'(0) y^n},
\end{equation*}
we obtain
\begin{equation*}
\L = -r^n-\s + (1-\a) \hat{y}(t) + \a \bs{\hat{y}(t)-\hat{y}(t+1)-\pi(t+1)}.
\end{equation*}

In conclusion, taking the log of the Euler equation \eqref{e:eulerdnl} yields
\begin{equation*}
-i^h(t) = -r^n-\s + (1-\a) \hat{y}(t) + \a \bs{\hat{y}(t)-\hat{y}(t+1)-\pi(t+1)}
\end{equation*}
Reshuffling the terms and noting that $i^h(t) = i(t) + \s$, we obtain the log-linearized Euler equation:
\begin{equation}
\hat{y}(t) = \a \hat{y}(t+1) - \bs{i(t) - r^n - \a \pi(t+1)}.
\label{e:eulerd}\end{equation}

\paragraph{Discounting.} Because $u'(0)>0$, we have
\begin{equation*}
\a = \frac{\b}{\b + u'(0) y^n} <1.
\end{equation*}
Thus, because the marginal utility of wealth is positive, the Euler equation is discounted: future output, $\hat{y}(t+1)$, appears discounted by the coefficient $\a<1$ in \eqref{e:eulerd}. Such discounting also appears in the presence of overlapping generations \cp{DGP12,EMR17}; heterogeneous agents facing borrowing constraints and cyclical income risk \cp{MNS17,AD18,B18}; consumers' bounded rationality \cp{G16}; incomplete information \cp{AL16}; bonds in the utility function \cp{CFJ17}; and borrowing costs increasing in household debt \cp{BP18}.

To make discounting more apparent, we solve the Euler equation forward:
\begin{equation*}
\hat{y}(t) = -\sum_{k=0}^{+\infty} \a^k \bs{i(t+k) - r^n - \a \pi(t+k+1)}.
\end{equation*}
The effect on current output of interest rates $k$ periods in the future is discounted by $\a^k<1$; hence, discounting is stronger for interest rates further in the future \cp[p.~821]{MNS17}.

\paragraph{Phillips curve.} Next we log-linearize the Phillips curve \eqref{e:phillipsdnl}. 

We start with the left-hand side of \eqref{e:phillipsdnl}. The first-order approximations of $x(x-1)$ and $\ln(x)$ around $x=1$ both are $x-1$. This means that up to second-order terms, we have $x(x-1)=\ln(x)$ around $x=1$. Hence, up to second-order terms, the following approximation holds around the natural steady state:
\begin{equation*}
\frac{p(t)}{p(t-1)}\bs{\frac{p(t)}{p(t-1)}-1}=\ln{\frac{p(t)}{p(t-1)}} = \pi(t).
\end{equation*}

We turn to the right-hand side of \eqref{e:phillipsdnl} and proceed similarly. We find that up to second-order terms, the following approximation holds around the natural steady state:
\begin{equation*}
\b \frac{p(t+1)}{p(t)}\bs{\frac{p(t+1)}{p(t)}-1} = \b\ln{\frac{p(t+1)}{p(t)}} = \b \pi(t+1).
\end{equation*}
Furthermore, \eqref{e:ytyn} implies that up to second-order terms, the ensuing approximation holds around the natural steady state:
\begin{equation*}
\frac{\e-1}{\g} \bs{\frac{y(t)}{y^n}-1} = \frac{\e-1}{\g} \hat{y}(t).
\end{equation*}

Combining all these results, we obtain the log-linearized Phillips curve:
\begin{equation}
\pi(t)= \b \pi(t+1) + \frac{\e-1}{\g} \hat{y}(t).
\label{e:phillipsd}\end{equation}

\section{Proofs}

We provide alternative proofs of propositions~1 and 2. These proofs are not graphical but algebraic; they are closer to the proofs found in the literature. We also complement the proof of proposition~4.

\subsection{Alternative proof of proposition~1}\label{a:classification}

We study the properties of the dynamical system generated by the Phillips curve (1) and Euler equation (4)  in normal times. The natural rate of interest is positive and monetary policy imposes $r(\pi) = r^n + (\f-1) \pi$.
	
\paragraph{Steady state.} A steady state $[y,\pi]$ must satisfy the steady-state Phillips curve (3) and steady-state Euler equation (7). These equations form a linear system:
\begin{align*}
\pi &=  \frac{\e\k}{\d\g a}  (y - y^n)\\
(\f-1) \pi &= - u'(0) (y - y^n).
\end{align*}
As $[y=y^n,\pi=0] $ satisfies both equations, it is a steady state. Furthermore the steady state is unique because the two equations are non-parallel. In the NK model, this is obvious since $u'(0)=0$. In the WUNK model, the slope of the second equation is $-u'(0)/(\f-1)$. If $\f>1$, the slope is negative. If $\f\in [0,1)$, the slope is positive and strictly greater than $u'(0)$ and thus than $\e\k/(\d\g a)$ (because (9) holds). In both cases, the two equations have different slopes.

\paragraph{Linearization.} The Euler-Phillips system is nonlinear, so we determine its properties by linearizing it around its steady state. We first write the Euler equation and Phillips curve as
\begin{align*}
\dot{y}(t)& = E(y(t),\pi(t)),\quad \text{where }E(y,\pi) = y [(\f-1) \pi + u'(0) (y - y^n)]\\
\dot{\pi}(t) & = P(y(t),\pi(t)),\quad \text{where }P(y,\pi) = \d \pi - \frac{\e\k}{\g a}(y - y^n).
\end{align*}
Around the natural steady state, the linearized Euler-Phillips system is
\begin{equation*}\bs{\begin{array}{c}
\dot{y}(t)\\
\dot{\pi}(t)
\end{array}}= \bs{\begin{array}{cc}
 \pd{E}{y} &  \pd{E}{\pi}\\ 
 \pd{P}{y}  &  \pd{P}{\pi}  \\ 
\end{array}} \bs{\begin{array}{c}
y(t)-y^n\\ 
\pi
\end{array}},\end{equation*}
where the derivatives are evaluated at $[y=y^n,\pi=0]$. We have 
\begin{align*}
\pd{E}{y} & = y^n u'(0),\qquad \pd{E}{\pi} = y^n (\f-1)\\
\pd{P}{y} & = - \frac{\e\k}{\g a},\qquad \pd{P}{\pi}=\d.
\end{align*}
Accordingly the linearized Euler-Phillips system is
\begin{equation}\bs{\begin{array}{c}
\dot{y}(t)\\
\dot{\pi}(t)
\end{array}}= \bs{\begin{array}{cc}
u'(0) y^n &   (\f-1) y^n \\ 
-\e\k/(\g a)  &  \d   \\ 
\end{array}} \bs{\begin{array}{c}
y(t)-y^n\\ 
\pi(t)
\end{array}}.\label{e:systemn}\end{equation}
We denote by $\M$ the matrix in \eqref{e:systemn}, and by $\m_1\in \C$ and $\m_2\in \C$ the two eigenvalues of $\M$, assumed to be distinct.
                                                                                                                                                                                                                                                                                                                                                                                                                                       
\paragraph{Solution with two real eigenvalues.} We begin by solving \eqref{e:systemn} when $\m_1$ and $\m_2$ are real and nonzero. Without loss of generality, we assume $\m_1<\m_2$. Then the solution takes the form
\begin{equation}
\bs{\begin{array}{c}
y(t)-y^n\\
 \pi(t)
\end{array}} 
=x_1 e^{\m_1 t} \v_1 + x_2 e^{\m_2 t} \v_2,
\label{e:solution}\end{equation}
where $\v_1 \in \R^2$ and $\v_2\in \R^2$ are the linearly independent eigenvectors respectively associated with the eigenvalues $\m_1$ and $\m_2$, and $x_1\in \R$ and $x_2\in \R$ are constants determined by the terminal condition \cp[p.~35]{HSD13}.

From \eqref{e:solution}, we see that the Euler-Phillips system is a source when $\m_1>0$ and $\m_2>0$. Moreover, the solutions are tangent to $\v_1$ when $t\to-\infty$ and are parallel to $\v_2$ when $t\to+\infty$. The system is a saddle when $\m_1<0$ and $\m_2>0$; in that case, the vector $\v_1$ gives the direction of the stable line (saddle path) while the vector $\v_2$ gives the direction of the unstable line. Lastly, when $\m_1<0$ and $\m_2<0$, the system is a sink. (See \inp[pp.~40--44]{HSD13}.)

\paragraph{Solution with two complex eigenvalues.} Next we solve \eqref{e:systemn} when $\m_1$ and $\m_2$ are complex conjugates. We write the eigenvalues as $\m_1=\t+i \vs$ and $\m_2=\t-i \vs$. We also write the eigenvector associated with $\m_1$ as $\v_1+i \v_2$, where the vectors $\v_1\in\R^2$ and $\v_2\in\R^2$ are linearly independent. Then the solution takes a more complicated form:
\begin{equation*}
\bs{\begin{array}{c}
y(t)-y^n\\
 \pi(t)
\end{array}} 
=e^{\t t}\bs{\v_1,\v_2}\bs{\begin{array}{cc}
\cos(\vs t) & \sin(\vs t)\\
-\sin(\vs t) & \cos(\vs t)
\end{array}} \bs{\begin{array}{c}
x_1\\
x_2
\end{array}},\end{equation*}
where $[\v_1,\v_2]\in \R^{2\times 2}$ is a $2\times 2$ matrix, and $x_1\in \R$ and $x_2\in \R$ are constants determined by the terminal condition \cp[pp.~44--55]{HSD13}.

These solutions wind periodically around the steady state, either moving toward it ($\t<0$) or away from it ($\t>0$). Hence, the Euler-Phillips system is a spiral source if $\t>0$ and a spiral sink if $\t<0$. In the special case $\t=0$, the solutions circle around the steady state: the Euler-Phillips system is a center. (See \inp[pp.~44--47]{HSD13}.)

\paragraph{Classification.} We classify the Euler-Phillips system from the trace and determinant of $\M$ \cp[pp.~61--64]{HSD13}. The classification relies on the property that $\tr(\M) = \m_1+\m_2$ and $\det(\M) = \m_1 \m_2$.
The following situations may occur in the NK and WUNK models:
\begin{itemize}
\item $\det(\M)<0$: Then the Euler-Phillips system is a saddle. This is because $\det(\M)<0$ indicates that $\m_1$ and $\m_2$ are real, nonzero, and of opposite sign. Indeed, if $\m_1$ and $\m_2$ were real and of the same sign, $\det(\M)=\m_1\m_2>0$; and if they were complex conjugates, $\det(\M)= \m_1 \ol{\m_1} = \Re(\m_1)^2+\Im(\m_1)^2>0$.
\item $\det(\M)>0$ and $\tr(\M)>0$: Then the Euler-Phillips system is a source. This is because $\det(\M)>0$ indicates that $\m_1$ and $\m_2$ are either real, nonzero, and of the same sign; or complex conjugates. Since in addition $\tr(\M)>0$,  $\m_1$ and $\m_2$ must be either real and positive, or complex with a positive real part. Indeed, if $\m_1$ and $\m_2$ were real and negative, $\tr(\M) = \m_1+\m_2<0$; if they were complex with a negative real part, $\tr(\M)=\m_1 + \ol{\m_1} = 2 \Re(\m_1) < 0$.
\end{itemize}

Using \eqref{e:systemn}, we compute the trace and determinant of $\M$:
\begin{align*}
\tr(\M) &= \d + u'(0) y^n\\
\det(\M) & = \d u'(0) y^n + (\f-1)\frac{\e\k}{\g a} y^n.
\end{align*} 

In the NK model, $u'(0)=0$, so $\tr(\M)=\d>0$ and 
\begin{equation*}
\det(\M) = (\f-1) \frac{y^n \e\k}{\g a}.
\end{equation*}
If $\f>1$, $\tr(\M)>0$ and $\det(\M)>0$, so the system is a source. If $\f<1$, $\det(\M)<0$, so the system is a saddle. 

In the WUNK model, $\tr(\M)>\d>0$. Further, using $\f-1\geq -1$ and (9), we have
\begin{equation*}
\det(\M) \geq  \d u'(0) y^n - \frac{\e\k}{\g a} y^n =  \d y^n \bs{u'(0) - \frac{\e\k}{\d\g a}} > 0. 
\end{equation*}
Since $\tr(\M)>0$ and $\det(\M)>0$, the system is a source.

\subsection{Alternative proof of proposition~2}

We study the properties of the dynamical system generated by the Phillips curve (1) and Euler equation (4)  at the ZLB. The natural rate of interest is negative and monetary policy imposes $r(\pi) = -\pi$.

\paragraph{Steady state.} A steady state $[y,\pi]$ must satisfy the steady-state Phillips curve (3) and the steady-state Euler equation (7). These equations form a linear system:
\begin{align}
\pi &=  \frac{\e\k}{\d\g a}  (y - y^n)\label{e:pcz}\\
\pi &= -r^n + u'(0) (y - y^n).\label{e:eez}
\end{align}
A solution to this system with positive output is a steady state.

In the NK model, $u'(0)=0$, so the system admits a unique solution:
\begin{align*}
\pi^z & = -r^n\\
y^z & =y^n-\frac{\d \g a}{\e\k}r^n.
\end{align*}
Since $r^n<0$, the solution satisfies $y^z>y^n>0$: the solution has positive output so it is a steady state. Hence the NK model admits a unique steady state at the ZLB: $[y^z,\pi^z]$, where $\pi^z>0$ (since $r^n<0$) and $y^z>y^n$. 

In the WUNK model, since (9) holds, the equations \eqref{e:pcz} and \eqref{e:eez} are non-parallel, so the system admits a unique solution, denoted $[y^z,\pi^z]$. Using \eqref{e:pcz} to substitute $y-y^n$ out of \eqref{e:eez}, we find that
\begin{equation}
\pi^z = \frac{r^n}{u'(0) \d \g a/(\e\k)-1}.
\label{e:piz}\end{equation}
Condition (9) implies that the denominator is positive. Since $r^n<0$, we conclude that $\pi^z<0$. 

Next, combining \eqref{e:pcz} and \eqref{e:piz}, we obtain
\begin{equation}
y^z =y^n +\frac{r^n}{u'(0)-\e\k/(\d\g a)}.
\label{e:yz}\end{equation}
Since (9) holds, the denominator of the fraction is positive. As $r^n<0$, we conclude that $y^z<y^n$.

Finally, to establish that $[y^z,\pi^z]$ is a steady state, we need to verify that $y^z>0$. According to \eqref{e:yz}, we need
\begin{equation*}
y^n > \frac{- r^n}{u'(0)-\e\k/(\d\g a)}.
\end{equation*}
Equations (5) and (9) indicate that 
\begin{equation*}
-r^n = u'(0)y^n-\d \quad\text{and}\quad u'(0)-\frac{\e\k}{\d\g a}>0.
\end{equation*}
The above inequality is therefore equivalent to
\begin{equation*}
\bs{u'(0)-\frac{\e\k}{\d\g a}} y^n > u'(0)y^n - \d.
\end{equation*}
Eliminating $u'(0)y^n$ on both sides, we find that this is equivalent to
\begin{equation*}
-\frac{\e\k y^n}{\d\g a}  > - \d,\quad\text{or}\quad \d^2>\frac{\e\k y^n}{\g a}.
\end{equation*}
Equation \eqref{e:yna} implies that
\begin{equation*}
\frac{\e\k y^n}{\g a} = \frac{\e-1}{\g}.
\end{equation*}
So we need to verify that 
\begin{equation*}
\d^2>\frac{\e-1}{\g}.
\end{equation*}
But we have imposed  $\d>\sqrt{(\e-1)/\g}$ in the WUNK model, to accommodate a positive natural rate of interest. We can therefore conclude that $y^z>0$, and that $[y^z,\pi^z]$ is a steady state.

\paragraph{Linearization.} The Euler-Phillips system is nonlinear, so we determine its properties by linearizing it. Around the ZLB steady state, the linearized Euler-Phillips system is 
\begin{equation}\bs{\begin{array}{c}
\dot{y}(t)\\
\dot{\pi}(t)
\end{array}}= \bs{\begin{array}{cc}
 u'(0) y^z &   -y^z \\ 
 -\e\k/(\g a)  &  \d   \\ 
\end{array}} \bs{\begin{array}{c}
y(t)-y^z\\ 
\pi(t)-\pi^z
\end{array}}.\label{e:systemz}\end{equation}
To obtain the matrix, denoted $\M$, we set $\f=0$ and replace $y^n$ by $y^z$ in the matrix from \eqref{e:systemn}.

\paragraph{Classification.} We classify the Euler-Phillips system~\eqref{e:systemz} by computing the trace and determinant of $\M$, as in appendix~\ref{a:classification}.  We have $\tr(\M)= \d+u'(0)y^z>0$ and 
\begin{equation*}
\det(\M) = \d y^z \bs{u'(0) - \frac{\e\k}{\d\g a}}.
\end{equation*} 
In the NK model, $u'(0)=0$ so $\det(\M)<0$, which implies that the Euler-Phillips system is a saddle. In the WUNK model, (9) implies that $\det(\M)>0$. Since in addition $\tr(\M)>0$, the Euler-Phillips system is a source. In fact, in the WUNK model, the discriminant of the characteristic equation of $\M$ is strictly positive:
\begin{align*}
\tr(\M)^2-4\det(\M) &= \d^2 + \bs{u'(0)y^n}^2+2 \d u'(0)y^n - 4\d u'(0) y^n +4\frac{\e\k}{\g a} y^n\\
										& = \bs{\d-u'(0)y^n}^2 +4\frac{\e\k}{\g a}y^n>0.
\end{align*}
Hence the eigenvalues of $\M$ are real, not complex: the Euler-Phillips system is a nodal source, not a spiral source.

\subsection{Complement to the proof of proposition~4}

We characterize the forward-guidance duration $\D^*$ for the NK model, and the ZLB duration $T^*$ for the WUNK model.

In the NK model, $\D^*$ is the duration of forward guidance that brings the economy on the unstable line of the ZLB phase diagram at time $T$ (figure~3, panel C). With longer forward guidance ($\D>\D^*$), the economy is above the unstable line at time $T$, and so it is connected to trajectories that come from the northeast quadrant of the ZLB phase diagram (figure~3, panel D). As a consequence, during ZLB and forward guidance, inflation is positive and output is above its natural level. Moreover, since the position of the economy at the end of the ZLB is unaffected by the duration of the ZLB, initial output and inflation become arbitrarily high as the ZLB duration of the ZLB approaches infinity.

In the WUNK model, for any forward-guidance duration, the economy at time $T$ is bound to be in the right-hand triangle of figure~4, panel D. All the points in that triangle are connected to trajectories that flow from the ZLB steady state, through the left-hand triangle of figure~4, panel D. For any of these trajectories, initial inflation $\pi(0)$ converges from above to the ZLB steady state's inflation $\pi^z$ as the ZLB duration $T$ goes to infinity. Since $\pi^z<0$, we infer that for each trajectory, there is a ZLB duration $\hat{T}$, such that for any $T>\hat{T}$, $\pi(0)<0$. Furthermore, as showed in panel D of figure~4, $y(0)<y^n$ whenever $\pi(0)<0$. The ZLB duration $T^*$ is constructed as $T^* = \max{\hat{T}}$. The maximum exists because the right-hand triangle is a closed and bounded subset of $\R^2$, so the set $\bc{\hat{T}}$ is a closed and bounded subset of $\R$, which admits a maximum. We know that the set $\bc{\hat{T}}$ is closed and bounded because the function that maps a position at time $T$ to a threshold $\hat{T}$ is continuous.

\section{Model with government spending}

We introduce government spending into the model of section~3. We compute the model's Euler equation and Phillips curve, linearize them, and use the linearized equations to construct the model's phase diagrams.

\subsection{Assumptions} 

We start from the model of section~3, and we assume that the government purchases a quantity $g_{j}(t)$ of each good $j\in[0,1]$. These quantities are aggregated into an index of public consumption
\begin{equation}
g(t) \equiv \bs{\int_{0}^{1} g_{j}(t)^{(\e-1)/\e}\,dj}^{\e/(\e-1)}.
\label{e:g}\end{equation}
Public consumption $g(t)$ enters separately into households' utility functions. Government expenditure is financed with lump-sum taxation.

Additionally, we assume that the disutility of labor is not linear but convex. Household~$j$ incurs disutility
\begin{equation*}
\frac{\k^{1+\eta}}{1+\eta} h_{j}(t)^{1+\eta}
\end{equation*}
from working, where $\eta> 0$ is the inverse of the Frisch elasticity. The utility function is altered to ensure that government spending affects inflation and private consumption.

\subsection{Euler equation \& Phillips curve}

We derive the Euler equation and Phillips curve just as in appendix~\ref{a:formal}.

The only new step is to compute the government's spending on each good. At any time $t$, the government chooses the amount $g_{j}(t)$ of each good $j\in[0,1]$ to minimize the expenditure 
 \begin{equation*}
 \int_0^1 p_{j}(t) g_{j}(t)\,dj
 \end{equation*}
subject to the constraint of providing an amount of public consumption $g$: 
\begin{equation*}
\bs{\int_{0}^{1} g_{j}(t)^{(\e-1)/\e}\,dj}^{\e/(\e-1)} = g(t).
\end{equation*}
To solve the government's problem at time $t$, we set up a Lagrangian:
\begin{equation*}
\Lc =   \int_0^1 p_{j}(t) g_{j}(t)\,dj + \Cc \cdot \bc{g - \bs{\int_{0}^{1} g_{j}(t)^{(\e-1)/\e}\,dj}^{\e/(\e-1)}},
\end{equation*}
where $\Cc$ is a Lagrange multiplier. We then follow the same steps as in the derivation of~\eqref{e:ydk}. The first-order conditions with respect to $g_{j}(t)$ for all $j\in [0,1]$ are $\pdx{\Lc}{g_j}=0$. These conditions imply
\begin{equation}
p_j(t) = \Cc \cdot \bs{\frac{g_j(t)}{g(t)}}^{-1/\e}.
\label{e:pj0}\end{equation} 
Appropriately integrating \eqref{e:pj0} over all $j\in [0,1]$, and using \eqref{e:p} and \eqref{e:g}, we find
\begin{equation}
\Cc =p(t).
\label{e:c}\end{equation}
Lastly, combining \eqref{e:pj0} and \eqref{e:c}, we obtain the government's demand for good $j$:
\begin{equation}
g_{j}(t)= \bs{\frac{p_j(t)}{p(t)}}^{-\e} g(t). 
\label{e:gj}\end{equation}

Next we solve household~$j$'s problem. We set up the current-value Hamiltonian:
\begin{align*}
\Hc_j &= \frac{\e}{\e-1} \ln{\int_{0}^{1} c_{jk}(t)^{(\e-1)/\e}\,dk} + u\of{\frac{b_{j}(t)-b(t)}{p(t)}} -\frac{1}{1+\eta} \bs{\frac{\k}{a}y^d_{j}(p_{j}(t),t)}^{1+\eta}-\frac{\g}{2} \pi_{j}(t)^2 \\
& + \Ac_{j}(t) \bs{i^h(t) b_{j}(t) + p_{j}(t)  y^d_{j}(p_{j}(t),t) - \int_0^1 p_{k}(t) c_{jk}(t)\,dk - \tau(t)}+\Bc_{j}(t) \pi_{j}(t) p_{j}(t).
\end{align*}

The terms featuring the consumption levels $c_{jk}(t)$ in the Hamiltonian are the same as in appendix~\ref{a:hamiltonian}, so the optimality conditions $\pdx{\Hc_j}{c_{jk}}=0$ remain the same. This implies that \eqref{e:aj0}, \eqref{e:aj}, and \eqref{e:cjk} remain valid. Adding the government's demand, given by \eqref{e:gj}, to households' demand, given by \eqref{e:cjk}, we obtain the total demand for good $j$ at time $t$:
\begin{equation*}
y^d_{j}(p_{j}(t),t) = g_{j}(t) +\int_0^1 c_{jk}(t) \, dk = \bs{\frac{p_{j}(t)}{p(t)}}^{-\e} y(t),
\end{equation*}
where $y(t)\equiv g(t) +\int_0^1 c_{j}(t)\,dj$ measures total consumption. The expression for $y^d_{j}(p_{j}(t),t)$ enters the Hamiltonian $\Hc_j$. 

The terms featuring the bond holdings $b_{j}(t)$ in the Hamiltonian are the same as in appendix~\ref{a:hamiltonian}. Therefore, the optimality condition $\pdx{\Hc_j}{b_{j}}=\d\Ac_{j}-\dot{\Ac}_{j}$ remains the same, and the Euler equation \eqref{e:eulerj} remains valid. In equilibrium, the Euler equation simplifies to
\begin{equation}
\frac{\dot{c}}{c}= r  - \d + \s + u'(0) c.
\label{e:eulerg}\end{equation}

The terms featuring inflation $\pi_{j}(t)$ in the Hamiltonian are also the same as in appendix~\ref{a:hamiltonian}. Thus, the optimality condition $\pdx{\Hc_j}{\pi_{j}}=0$ is unchanged, and \eqref{e:bj} and \eqref{e:bjdot} hold.

Last, because the disutility from labor is convex, the optimality condition $\pdx{\Hc_j}{p_{j}}=\d \Bc_{j}-\dot{\Bc}_{j}$ is modified. The condition now gives
\begin{equation*}
\frac{\e}{p_j} \bp{\frac{\k}{a}y_{j}}^{1+\eta}+ (1-\e) \Ac_{j} y_{j} + \Bc_{j} \pi_{j} = \d \Bc_{j}-\dot{\Bc}_{j},
\end{equation*}
which can be rewritten
\begin{equation*}
\pi_{j} - \frac{(\e-1) y_{j} \Ac_{j}}{\Bc_{j} p_{j}} \bs{p_{j} -\frac{\e}{\e-1} \bp{\frac{\k}{a}}^{1+\eta}\frac{y_{j}^{\eta}}{\Ac_{j}}} = \d-\frac{\dot{\Bc}_{j}}{\Bc_{j}}.
\end{equation*}
Combining this equation with \eqref{e:aj}, \eqref{e:bj}, and \eqref{e:bjdot}, we obtain the household's Phillips curve:
\begin{equation}
\frac{\dot{\pi}_{j}}{\pi_{j}} = \d + \frac{(\e-1) y_{j}}{\g c_{j} \pi_{j}} \bs{\frac{p_{j}}{p}-\frac{\e}{\e-1}\bp{\frac{\k}{a}}^{1+\eta}y_{j}^{\eta}c_{j}}.
\label{e:phillipsjg}\end{equation}
In equilibrium, the Phillips curve simplifies to
\begin{equation}
\dot{\pi} = \d \pi+ \frac{(\e-1) (c+g)}{\g c} \bs{1-\frac{\e}{\e-1}\bp{\frac{\k}{a}}^{1+\eta}(c+g)^{\eta} c},
\label{e:phillipsg}
\end{equation}
where $c+g = y$ is aggregate output.

\subsection{Linearized Euler-Phillips system}

We now linearize the Euler-Phillips system around the natural steady state, which has zero inflation and no government spending. The analysis of the model with government spending is based on this linearized system.

Since $\dot{\pi} = \pi = g =0 $ at the natural steady state, \eqref{e:phillipsg} implies that the natural level of consumption is 
\begin{equation*}
c^n = \bp{\frac{\e-1}{\e}}^{1/(1+\eta)} \frac{a}{\k}.
\end{equation*}
Since $\dot{c} = 0$ and $c=c^n$ at the natural steady state, \eqref{e:eulerg} implies that the natural rate of interest is 
\begin{equation*}
r^n = \d-\s - u'(0)c^n. 
\end{equation*}

\paragraph{Euler equation.} We first linearize the Euler equation \eqref{e:eulerg} around the point $[c=c^n,\pi=0]$. We consider two different monetary-policy rules. First, when monetary policy is normal, $r(\pi) = r^n + \bp{\f-1} \pi$. Then the Euler equation is $\dot{c}=E(c,\pi)$, where 
\begin{equation*}
E(c,\pi) = c \bs{(\f-1) \pi + u'(0) (c - c^n)}.
\end{equation*} 
The linearized version is 
\begin{equation*}
\dot{c} = E(c^n,0)+ \pd{E}{c}(c-c^n)+\pd{E}{\pi} \pi,
\end{equation*}
where the derivatives are evaluated at $[c=c^n,\pi=0]$. We have
\begin{equation*}
E(c^n,0)= 0,\qquad  \pd{E}{c}= c^n u'(0),\qquad \pd{E}{\pi}= c^n (\f-1).
\end{equation*}
So the linearized Euler equation with normal monetary policy is 
\begin{equation}
\dot{c} = c^n \bs{(\f-1) \pi + u'(0)(c-c^n)}.
\label{e:eulergln}\end{equation}

Second, when monetary policy is at the ZLB, $r(\pi)=-\pi$. Then the Euler equation can be written $\dot{c}=E(c,\pi)$ where
\begin{equation*}
E(c,\pi) = c \bs{-r^n-\pi + u'(0) (c - c^n)}.
\end{equation*} 
The linearized version is 
\begin{equation*}
\dot{c} = E(c^n,0) + \pd{E}{c}(c-c^n)+ \pd{E}{\pi} \pi ,
\end{equation*}
where the derivatives are evaluated at $[c=c^n,\pi=0]$. We have
\begin{equation*}
E(c^n,0)=-c^n r^n, \qquad \pd{E}{c}= c^n u'(0), \qquad \pd{E}{\pi}= -c^n.
\end{equation*}
So the linearized Euler equation at the ZLB is 
\begin{equation}
\dot{c} = c^n \bs{-r^n - \pi + u'(0)(c-c^n)}.
\label{e:eulerglz}\end{equation}
In steady state, at the ZLB, the linearized Euler equation becomes
\begin{equation}
\pi = -r^n +u'(0) (c-c^n).
\label{e:eulerglssz}\end{equation}

\paragraph{Phillips curve.} Next we linearize the Phillips curve \eqref{e:phillipsg} around the point $[c=c^n,\pi=0,g=0]$. The Phillips curve can be written $\dot{\pi} = P(c,\pi,g)$ where 
\begin{equation*}
P(c,\pi,g) = \d \pi+ \frac{(\e-1) (c+g)}{\g c} \bs{1-\frac{\e}{\e-1}\bp{\frac{\k}{a}}^{1+\eta}(c+g)^{\eta} c}.
\end{equation*}
The linearized version is 
\begin{equation*}
\dot{\pi} = P(c^n,0,0)+ \pd{P}{c}(c-c^n)+\pd{P}{\pi} \pi +  \pd{P}{g}g,
\end{equation*}
where the derivatives are evaluated at $[c=c^n,\pi=0,g=0]$. We have 
\begin{align*}
P(c^n,0,0)& = 0\\
\pd{P}{c}&= - \frac{\e}{\g}\bp{\frac{\k}{a}}^{1+\eta}(1+\eta) \bp{c^n}^\eta = - (1+\eta) \frac{\e\k}{\g a} \bp{\frac{\e-1}{\e}}^{\eta/(1+\eta)}\\
\pd{P}{\pi}&= \d\\
\pd{P}{g}&=- \frac{\e}{\g}\bp{\frac{\k}{a}}^{1+\eta} \eta \bp{c^n}^\eta = - \eta \frac{\e\k}{\g a} \bp{\frac{\e-1}{\e}}^{\eta/(1+\eta)}.
\end{align*}
Hence, the linearized Phillips curve is 
\begin{equation}
\dot{\pi} = \d \pi -  \frac{\e\k}{\g a} \bp{\frac{\e-1}{\e}}^{\eta/(1+\eta)}\bs{(1+\eta) \bp{c-c^n} + \eta g}.
\label{e:phillipsgl}\end{equation}
In steady state, the linearized Phillips curve becomes
\begin{equation}
\pi =  - \frac{\e\k}{\d \g a} \bp{\frac{\e-1}{\e}}^{\eta/(1+\eta)}\bs{(1+\eta) (c-c^n) + \eta g}.
\label{e:phillipsglss}\end{equation}

\subsection{Phase diagrams}

Using the linearized Euler-Phillips system, we construct the phase diagrams of the NK and WUNK models with government spending.

\begin{figure}[p]
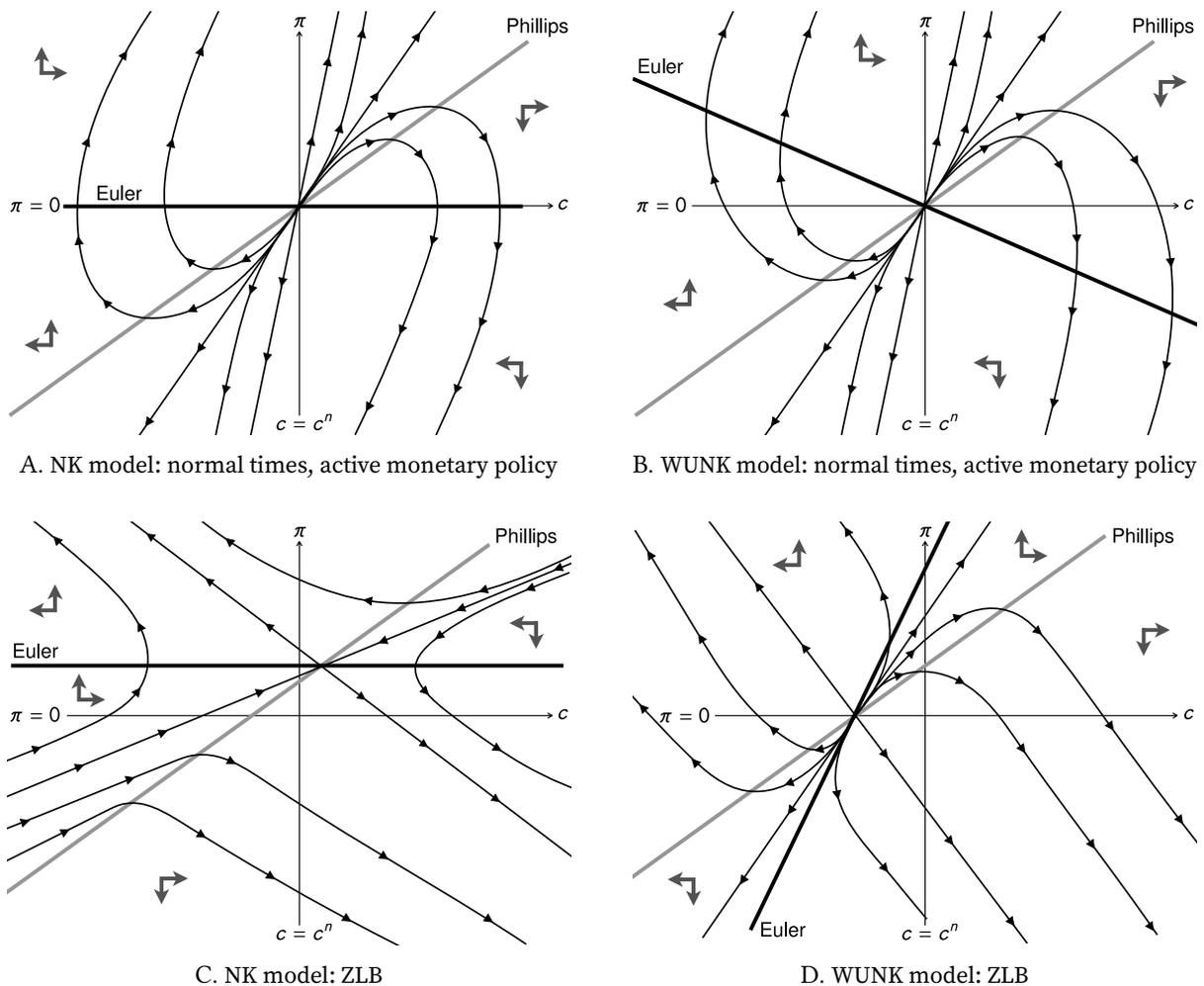

\subcaptionbox{NK model: normal times, active monetary policy}{\includegraphics[scale=0.21,page=29]{\pdf}}\hfill
\subcaptionbox{WUNK model: normal times, active monetary policy}{\includegraphics[scale=0.21,page=31]{\pdf}}\vspace{0.5cm}
\subcaptionbox{NK model: ZLB}{\includegraphics[scale=0.21,page=30]{\pdf}}\hfill
\subcaptionbox{WUNK model: ZLB}{\includegraphics[scale=0.21,page=32]{\pdf}}
\caption{Phase diagrams of the linearized Euler-Phillips system in the NK and WUNK models with government spending}
\note{The figure displays phase diagrams for the linearized Euler-Phillips system in the model with government spending: $c$ is private consumption; $\pi$ is inflation; $c^n$ is the natural level of consumption; the Euler line is the locus $\dot{c}=0$; the Phillips line is the locus $\dot{\pi}=0$; the trajectories are solutions to the system, plotted for $t$ going from $-\infty$ to $+\infty$. The four panels contrast various cases. The NK model is the standard New Keynesian model. The WUNK model is the same model, except that the marginal utility of wealth is not zero but is sufficiently large to satisfy (12). In normal times with active monetary policy, the natural rate of interest $r^n$ is positive, the monetary-policy rate is given by $i=r^n+\f\pi$ with $\f>1$, and government spending is zero; the Euler-Phillips system is composed of \eqref{e:eulergln} with $\f>1$ and \eqref{e:phillipsgl} with $g=0$. At the ZLB, the natural rate of interest is negative, the monetary-policy rate is zero, and government spending is positive; the Euler-Phillips system is composed of \eqref{e:eulerglz} and \eqref{e:phillipsgl} with $g>0$. The figure shows that in the NK model, the Euler-Phillips system is a source in normal times with active monetary policy (panel~A); but the system is a saddle at the ZLB (panel~C). In the WUNK model, by contrast, the Euler-Phillips system is a source both in normal times and at the ZLB (panels~B and D). (Panels~A and B display a nodal source, but the system could also be a spiral source, depending on the value of $\f$; in panel~D the system is always a nodal source.)}
\label{f:phasea}\end{figure}

\paragraph{Normal times.} We first construct the phase diagrams for normal times with active monetary policy. The linearized Euler-Phillips system is composed of \eqref{e:eulergln} with $\f>1$ and \eqref{e:phillipsgl} with $g=0$.

We construct a phase diagram with private consumption $c$ on the horizontal axis and inflation $\pi$ on the vertical axis. We follow the methodology developed in section~3: we plot the loci $\dot{\pi}=0$ and $\dot{c}=0$, and then determine the sign of $\dot{\pi}$ and $\dot{c}$ in the four quadrants of the plan delimited by the two loci. The resulting phase diagrams are displayed in panels A and B of figure~\ref{f:phasea}. They are similar to the phase diagrams in the basic model (panels A and B of figure~1).\footnote{The phase diagrams of figure~1 have output $y$ on the horizontal axis instead of private consumption $c$. But $y=c$ in the basic model (government spending is zero), so phase diagrams with $c$ on the horizontal axis would be identical.}

The phase diagrams show that in normal times, with active monetary policy, the Euler-Phillips system is a source in the NK and WUNK models. An algebraic approach confirms this result. The linearized Euler-Phillips system can be written
\begin{equation*}
\bs{\begin{array}{c}
\dot{c}\\
\dot{\pi}
\end{array}}= \bs{\begin{array}{cc}
 u'(0) c^n &   (\f-1) c^n \\ 
- (1+\eta)\frac{\e\k}{\g a}\bp{\frac{\e-1}{\e}}^{\eta/(1+\eta)}  &  \d \\
\end{array}} \bs{\begin{array}{c}
c-c^n\\ 
\pi
\end{array}}.
\end{equation*}
We denote the above matrix by $\M$. We classify the Euler-Phillips system using the trace and determinant of $\M$, as in appendix~\ref{a:classification}:
\begin{align*}
\tr(\M) &= \d + u'(0) c^n\\
\det(\M) & = \d c^n \bs{u'(0)  + (\f-1)(1+\eta) \frac{\e\k}{\d\g a}\bp{\frac{\e-1}{\e}}^{\eta/(1+\eta)}}.
\end{align*} 

In the NK model, $u'(0)=0$ so $\tr(\M)=\d>0$ and the sign of $\det(\M)$ is given by the sign of $\f-1$. Accordingly when monetary policy is active ($\f>1$), $\det(\M)>0$: the Euler-Phillips system is a source. In contrast, when monetary policy is passive ($\f<1$), $\det(\M)<0$: the Euler-Phillips system is a saddle.

In the WUNK model, $\tr(\M)>\d>0$. Moreover, $\f-1\geq -1$ for any $\f\geq 0$, so we have
\begin{equation*}
\det(\M) \geq \d c^n \bs{u'(0) - (1+\eta)\frac{\e\k}{\d\g a}\bp{\frac{\e-1}{\e}}^{\eta/(1+\eta)}}.
\end{equation*}
The WUNK assumption (12) says that the term in square brackets is positive, so $\det(\M)>0$. We conclude that the Euler-Phillips system is a source whether monetary policy is active or passive.

\paragraph{ZLB.} We turn to the phase diagrams at the ZLB. The linearized Euler-Phillips system is composed of \eqref{e:eulerglz} and \eqref{e:phillipsgl} with $g>0$.

Once again, we follow the methodology developed in section~3 to construct the phase diagrams. The resulting phase diagrams are displayed in panels~C and D of figure~\ref{f:phasea}. The diagrams have the same properties as in the basic model (panels C and D of figure~1), but for one difference: the Phillips line shifts upward because government spending is positive. Hence, the Phillips line lies above the point  $[c=c^n,\pi=0]$. While this shift does not affect the classification of the Euler-Phillips system (source or saddle), it changes the location of the steady state. In fact, by solving the system given by \eqref{e:eulerglssz} and \eqref{e:phillipsglss}, we find that private consumption and inflation at the ZLB steady state are
\begin{align}
c^g &=c^n + \frac{r^n+ \frac{\e\k}{\d\g a}\bp{\frac{\e-1}{\e}}^{\eta/(1+\eta)} \eta g }{u'(0)-(1+\eta) \frac{\e\k}{\d\g a}\bp{\frac{\e-1}{\e}}^{\eta/(1+\eta)}}\label{e:yzg}\\
\pi^g &= \frac{(1+\eta) r^n +u'(0) \eta g}{u'(0)\frac{\d\g a}{\e\k}\bp{\frac{\e}{\e-1}}^{\eta/(1+\eta)}-(1+\eta)}.\label{e:pizg}
\end{align}
Steady-state consumption may be above or below natural consumption, depending on the amount of government spending. In the WUNK model, inflation may be positive or negative, depending on the amount of government spending. 

The phase diagrams show that at the ZLB, the Euler-Phillips system is a source in the WUNK model but a saddle in the NK model. An algebraic approach confirms this classification. Rewritten in canonical form, the linearized Euler-Phillips system becomes
\begin{equation*}
\bs{\begin{array}{c}
\dot{c}\\
\dot{\pi}
\end{array}} 
=
\bs{\begin{array}{cc}
 u'(0) c^n &   -c^n \\ 
-  (1+\eta) \frac{\e\k}{\g a}\bp{\frac{\e-1}{\e}}^{\eta/(1+\eta)}  &  \d   \\ 
\end{array}}
\bs{\begin{array}{c}
c-c^g\\ 
\pi-\pi^g
\end{array}}.
\end{equation*}
We denote the above matrix by $\M$. We classify the Euler-Phillips system using the trace and determinant of $\M$, as in appendix~\ref{a:classification}:
\begin{align*}
\tr(\M) &= \d + u'(0) c^n\\
\det(\M) & = \d c^n \bs{u'(0)  - (1+\eta) \frac{\e\k}{\d\g a}\bp{\frac{\e-1}{\e}}^{\eta/(1+\eta)}}.
\end{align*} 
In the NK model, $u'(0)=0$ so $\det(\M)<0$, indicating that the Euler-Phillips system is a saddle. In the WUNK model, condition (12) implies that $\det(\M)>0$; since we also have $\tr(\M)>0$, we conclude that the Euler-Phillips system is a source. We can also show that $\tr(\M)^2-4\det(\M)>0$, which indicates that the system is a nodal source, not a spiral source.

\section{Proofs with government spending}

We complement the proofs of propositions 5 and 9, which pertain to the model with government spending.

\subsection{Complement to the proof of proposition~5}

We characterize the amount $g^*$ in the NK model, and we compute the limit of the government-spending multiplier in the WUNK model.

In the NK model, the amount $g^*$ of government spending is the amount that makes the unstable line of the dynamical system go through the natural steady state. With less spending than $g^*$ (panel B of figure~5), the natural steady state is below the unstable line and is connected to trajectories coming from the southwest quadrant of the phase diagram. Hence, for $g<g^*$, $\lim_{T\to \infty}c(0;g)=-\infty$. With more spending than $g^*$ (panel D of figure~5), the natural steady state is above the unstable line and is connected to trajectories coming from the northeast quadrant. Hence, for $g>g^*$, $\lim_{T\to \infty}c(0;g)=+\infty$. Accordingly, for any $s>0$, $\lim_{T\to \infty}m(g^*,s)=+\infty$.   

In the WUNK model, when the ZLB is infinitely long-lasting, the economy jumps to the ZLB steady state at time $0$: $\lim_{T\to \infty}c(0;g)=c^g(g)$, where $c^g(g)$ is given by \eqref{e:yzg}. The steady-state consumption $c^g(g)$ is linear in government spending $g$, with a coefficient in front of $g$ of
\begin{equation*}
\frac{\eta}{u'(0)\frac{\d\g a}{\e\k}\bp{\frac{\e}{\e-1}}^{\eta/(1+\eta)}-(1+\eta)}.
\end{equation*}
Accordingly, for any $s>0$, we have
\begin{align*}
\lim_{T\to \infty}m(g,s)&=1+\frac{\lim_{T\to \infty}c(0;g+s/2)-\lim_{T\to \infty}c(0;g-s/2)}{s}\\
&=1+\frac{c^g(g+s/2)-c^g(g-s/2)}{s}\\
&= 1+\frac{\eta}{u'(0)\frac{\d\g a}{\e\k}\bp{\frac{\e}{\e-1}}^{\eta/(1+\eta)}-(1+\eta)},
\end{align*}
which corresponds to (13).

\subsection{Complement to the proof of proposition~9}

We compute the government-spending multiplier at the ZLB in the WUNK model. Private consumption and inflation at the ZLB steady state are determined by \eqref{e:yzg} and \eqref{e:pizg}. The coefficients in front of government spending $g$ in these expressions are
\begin{equation*}
\frac{\eta}{u'(0)\frac{\d\g a}{\e\k}\bp{\frac{\e}{\e-1}}^{\eta/(1+\eta)}-(1+\eta)}\quad\text{and}\quad\frac{u'(0) \eta}{u'(0)\frac{\d\g a}{\e\k}\bp{\frac{\e}{\e-1}}^{\eta/(1+\eta)}-(1+\eta)}.
\end{equation*}
Since (12) holds, both coefficients are positive. Hence, an increase in $g$ raises private consumption and inflation. Moreover, $dc/dg$ is given by the first of these coefficient, which immediately yields the expression for the multiplier $dy/dg=1+dc/dg$.

\section{WUNK assumption in terms of estimable statistics}

We re-express the WUNK assumption in terms of estimable statistics. We first work on the model with linear disutility of labor, in which the assumption is given by (9). We then turn to the model with convex disutility of labor, in which the assumption is given by (12).

\subsection{Linear disutility of labor}

When the disutility of labor is linear, the WUNK assumption is given by (9). Multiplying (9) by $y^n$, we obtain
\begin{equation*}
u'(0) y^n > \frac{1}{\d}\cdot\frac{y^n \e\k}{\g a}.
\end{equation*}
The time discount rate $\d$ has been estimated in numerous studies. We therefore only need to express $u'(0) y^n$ and $y^n \e\k/(\g a)$ in terms of estimable statistics.

First, the definition of the natural rate of interest, given by (5), implies that $u'(0) y^n = \d - \s - r^n$. Following the New Keynesian literature \eg[p.~20]{W11}, we set the financial-intermediation spread to $\s=0$ in normal times. Hence, in normal times,  $u'(0) y^n = \d - r^n$. Thus, $u'(0) y^n$ can be measured from the gap between the discount rate $\d$ and the average natural rate of interest $r^n$---both of which have been estimated by many studies. 

Second, we show that $y^n \e\k/(\g a)$ can be measured from estimates of the New Keynesian Phillips curve. To establish this, we compute the discrete-time New Keynesian Phillips curve arising from our continuous-time model. We start from the first-order approximation 
\begin{equation*}
\pi(t) = \pi(t+dt) - \dot{\pi}(t+dt) dt
\end{equation*}
and use (1) to measure $\dot{\pi}(t+dt)$. We obtain
\begin{equation*}
\pi(t) = \pi(t+dt) - \d \pi(t+dt) dt + \frac{y^n \e\k}{\g a}  \cdot \frac{y(t) - y^n}{y^n} dt.
\end{equation*}
(We have replaced $y(t+dt)dt$ by $y(t)dt$ since the difference between the two is of second order.) Setting the unit of time to one quarter (as in the empirical literature) and $dt=1$, we obtain 
\begin{equation}
\pi(t) = (1-\d)\pi(t+1) + \frac{y^n \e\k}{\g a}  x(t),
\label{e:mps14}\end{equation}
where $\pi(t)$ is quarterly inflation at time $t$, $\pi(t+1)$ is quarterly inflation at time $t+1$, and 
\begin{equation*}
x(t) = \frac{y(t) - y^n}{y^n}
\end{equation*}
is the output gap at time $t$. Equation \eqref{e:mps14} is a typical New Keynesian Phillips curve, so we can measure $y^n \e\k/(\g a)$ by estimating the coefficient on output gap in a standard New Keynesian Phillips curve---which has been done many times.

To sum up, we rewrite the WUNK assumption as 
\begin{equation*}
\d-r^n > \frac{\l}{\d},
\end{equation*} 
where $\d$ is the time discount rate, $r^n$ is the average natural interest rate, and $\l$ is the output-gap coefficient in a standard New Keynesian Phillips curve. This is just (14). 

\subsection{Convex disutility of labor}

When the disutility of labor is convex, the WUNK assumption is given by (12):
\begin{equation*}
u'(0)y^n > \frac{1}{\d}\cdot \frac{y^n \e\k}{\g a}\bp{\frac{\e-1}{\e}}^{\eta/(1+\eta)}(1+\eta).
\end{equation*}
To rewrite this condition in terms of estimating statistics, we follow the previous method. The only change occurs when computing the discrete-time New Keynesian Phillips curve arising from the continuous-time model. To measure $\dot{\pi}(t+dt)$, we use \eqref{e:phillipsgl} with $g=0$ and $c=y$. As a result, \eqref{e:mps14} becomes 
\begin{equation*}
\pi(t) = (1-\d)\pi(t+1) + \frac{y^n \e\k}{\g a}\bp{\frac{\e-1}{\e}}^{\eta/(1+\eta)}(1+\eta)  x(t),
\end{equation*}
where $\pi(t)$ and $\pi(t+1)$ are quarterly inflation rates and $x(t)$ is the output gap. This is just a typical New Keynesian Phillips curve. Hence, again, we can measure
\begin{equation*}
\frac{y^n \e\k}{\g a}\bp{\frac{\e-1}{\e}}^{\eta/(1+\eta)}(1+\eta)
\end{equation*}
by estimating the output-gap coefficient in a standard New Keynesian Phillips curve.

To conclude, just as with a linear disutility of labor, we can rewrite the WUNK assumption as 
\begin{equation*}
\d-r^n > \frac{\l}{\d},
\end{equation*} 
where $\d$ is the time discount rate, $r^n$ is the average natural rate of interest, and $\l$ is the output-gap coefficient in a standard New Keynesian Phillips curve.

\end{document}